\newif\ifrevtex
\revtextrue

\newif\ifmoneyagg
\moneyaggtrue

\newif\ifsfb
\sfbfalse

\newif\ifanonymous
\anonymousfalse

\newif\iftitlepage
\titlepagefalse

\ifrevtex
\documentclass[onecolumn,rmp,tightenlines,nofootinbib,showkeys,longbibliography,11pt]{revtex4}
\else
\documentclass[onecolumn,authoryear,3p,10pt]{elsarticleCJE}
\fi

\ifrevtex
\else
\usepackage{setspace}
\fi

\usepackage{graphicx}
\usepackage{color}
\usepackage{dsfont}
\usepackage{hyperref}
\usepackage{eurosym}
\usepackage{textcomp}
\usepackage{multirow}
\usepackage{amsmath}

\newcommand{\be}{\begin{equation}}
\newcommand{\ee}{\end{equation}}
\newcommand{\bea}{\begin{eqnarray}}
\newcommand{\eea}{\end{eqnarray}}

\newcommand{\CJE}[1]{#1}

\begin{document}

\title{Graph representation of balance sheets: from exogenous to endogenous money}

\ifanonymous
\else
%\title{Endogenous money and state consolidation from graphs of balance sheets}
\author{Cyril Pitrou}
\fi

%For revtex
\ifrevtex

\ifanonymous
\else
\email{cyril.pitrou@upmc.fr}
\affiliation{Universit\'e Pierre~\&~Marie Curie - Paris VI, Sorbonne
  Universit\'es, Institut d'Astrophysique de Paris, CNRS-UMR 7095, 98 bis Bd Arago, 75014 Paris, France}
\fi


\else
%For Elsevier
\ifanonymous
\else
\ead{cyril.pitrou@upmc.fr}
\address{Universit\'e Pierre~\&~Marie Curie - Paris VI, Sorbonne
  Universit\'es, Institut d'Astrophysique de Paris, CNRS-UMR 7095, 98 bis Bd
  Arago, 75014 Paris, France}
\fi

\fi

%\date{April 14, 2015}
\date{\today}

%%%%%%%%%%%%%%%%%%%%%%%%%%%%%%%%%%%%%%%%%%%%%%%%%%%%%%%%%%%%%%%%%%%%%%%
\begin{abstract}
The nature of monetary arrangements is often discussed without any
reference to its detailed construction. We present a graph representation \CJE{that} allows for a clear understanding of modern
monetary systems. First, we show that systems based on commodity
money are incompatible with credit. We then study the current chartalist systems based on pure fiat
money, and we discuss the consolidation of the central bank with the
Treasury. We obtain a visual explanation about how commercial banks are responsible for endogenous
money creation whereas the Treasury and the central bank are in charge
of the total amount of net money.
Finally we draw an analogy between systems based on gold
convertibility and currency pegs to show that fixed exchange rates can never be maintained.
\iftitlepage
\\
\newline
\footnotesize{The author thank D. D\'efossez-Carme, L. Gastard,
  M. Lavoie, E. Le H\'eron, S. Renaux-Petel, E. Rolet and  J.-P. Uzan
  for discussions and comments, and Johanna Skrzypczyk for assistance with editing the manuscript.}
\else
\fi
\end{abstract}
\maketitle

\iftitlepage
\end{document}
\else
\fi

%%%%%%%%%%%%%%%%%%%%%%%%%%%%%%%%%%%%%%%%%%%%%%%%%%%%%%%%%%%%%%%%%%%%%%%

\section*{Introduction}

Understanding the fundamental nature of money is paramount for macroeconomic models. Indeed, over the past
century the description of money was central to the building of
simplified macroeconomic models, and monetary economics became a
field of macroeconomics in itself. The {\it classicals} first
postulated that money was neutral and avoided the need for any
suitable description. The {\it new classicals}, and most prominently
the monetarists reached a more nuanced version with the quantity theory of money,
stating that money was neutral in the long run. Again, the debate was
not on the structure of the monetary system but rather on its
implications for macroeconomic theories. The debate between
neoclassicals and Keynesians led instead to theories of money being
\CJE{addressed as} theories of interest rates.

There is certainly no unique theory of money, since money is not a
universal concept. Indeed, any monetary arrangement in a society is different and leads to a specific description. We should describe the various possibilities of {\it  monetary systems} rather
than talk about a unique {\it theory of money}, including those which have
been or are realized, but also allow for the description of possible
systems not yet realized in practice, as required by the general
democratic debate. This is nicely summarized by~\citet[p1]{Minsky1996a}: {\it ``[\dots] relevant theory is not a
  compendium of propositions derived from axioms assumed to be
  universally true: economic theory is not a subdivision of mathematics. Relevant theory is the result of the exercise of imagination and logical powers on observations that are due to
experience: it yields propositions about the operation of an actual economy"} . 

Monetary systems can be \CJE{divided} in two major categories: metallism and
chartalism. Oversimplifying, one can say that metallism is the description of monetary systems in which money
is backed by a real asset, often gold, whereas chartalism is the
description of monetary systems based on pure fiat money that
originates from the state's ability to raise taxes (see \citet{StephanieBell2001} for a historical perspective and~\citet{Innes1913,Lerner,Knapp} for seminal papers). Since the
end of the Bretton-Woods system in 1971, which was already a remote form
of convertibility, this latter descriptive framework has received renewed
attention and is often named neo-chartalism in that respect~\citep{Wray1990}.

In this article, we emphasize that all monetary arrangements are
  intrinsically weighted graphs, because accounting is
performed with double entry balance sheets. \CJE{As Minsky explained} \citep[p12]{Minsky1992a}: {\it ``A capitalist economy  can be described by a set of interrelated balance sheets and income
  statements''}, implying that graphs are the natural theoretical tool to
describe monetary systems. \CJE{Graphs} are better suited for
\CJE{representing} the global structure of financial systems than
lists of individual balance sheets, since they allow for a clear visualisation of the relations between members of the system. Such graphs are rather common in
representations of flows in macroeconomics, but are generally absent
\CJE{from} the description of stocks, although they have already been extensively
used to study the systemic risks in networks of commercial banks~[see
e.g.~\citet{SheldonMaurer,UpperWorms,Upper2011,Gai,Wells2004}]. 
In this article, we argue that graphs are the natural language in which discussions
about the nature of monetary systems should be \CJE{spoken}. By
representing a complex structure in a very compact form, \CJE{graphs depict
information clearly}, just like a very deep equation
carries a powerful meaning in just one line, and this allows for
clearer representations of the global architecture of monetary
systems. Weighted graphs \CJE{open the door to discussion of} major
topics associated with monetary systems, as they allow for an unambiguous and visual representation of the monetary arrangements.

In a physical system with a high number of particles, there are complex structures
\CJE{that} emerge from simple local laws, and they are quite often
very difficult to understand. Similarly, the financial relations
between all actors of a monetary system are complex and simple at the
same time. They are simple because double-entry bookkeeping is
understood at all levels of the system, but there are complex
financial structures which emerge out of the simple local laws of
accounting, due to a high number of financial interactions. As a
result, \CJE{if the inputs of a graph are understood by all}, the different
schools of economic thought should agree on the graph representation of any monetary
structure. Indeed, we show that each given financial situation can be
summarized by drawing the corresponding graph, and this is
  crucial to discuss controversial topics. In particular, we
illustrate {\it i)} how public deficits is the source of money
creation, {\it ii)} that gold convertibility and currency boards are
inherently incompatible with credit, and {\it iii)} that credits do not
originate from deposits, all these subjects allowing to discuss further the
nature of money. \CJE{Disagreements therefore lie} in the interpretation, that is on
the words we drape around a given graph. We will argue that these
differences of interpretation are related to the various possible consolidations which can be made in a monetary system, when trying to grasp the complex emerging structures.\\

%This article is organized as follows. We first present the building
%blocks of our graph formalism in \S~\ref{SecGenForm}. We then describe
%the endogenous money creation by commercial banks in \S~\ref{SecEndo}
%and turn to the role of the state in \S~\ref{SecState}. We
%discuss briefly the interrelation of various currencies in
%\S~\ref{SecFor} and show that currency pegs can never be maintained.

%DONE

\section{General formalism for graphs of monetary arrangements}\label{SecGenForm}

\subsection{Commodities and units of account}

A general monetary arrangement is a collection of debts
  between various members of the system. However, before describing how this
  system of debts is organized,  \CJE{one} must first specify what is owed in
  these debts, and we thus need an appropriate description for ownership. The simplest
  \CJE{item that} can be possessed and owed is a commodity and in that case, the unit of account (UoA) is any natural unit for that specific commodity, e.g. a unit of mass for gold.

\begin{figure}[!htb]
 \includegraphics[height=0.32\textwidth]{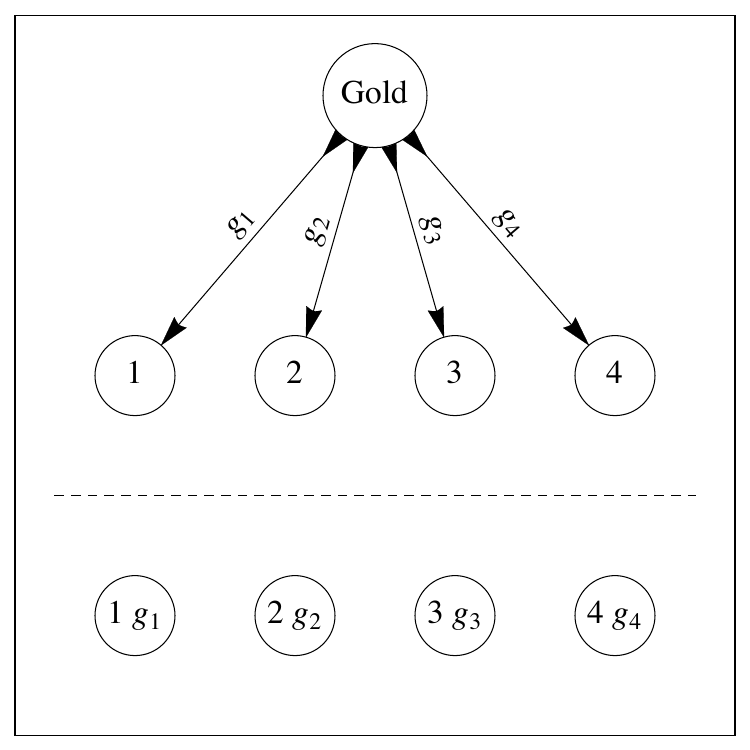}
 \includegraphics[height=0.32\textwidth]{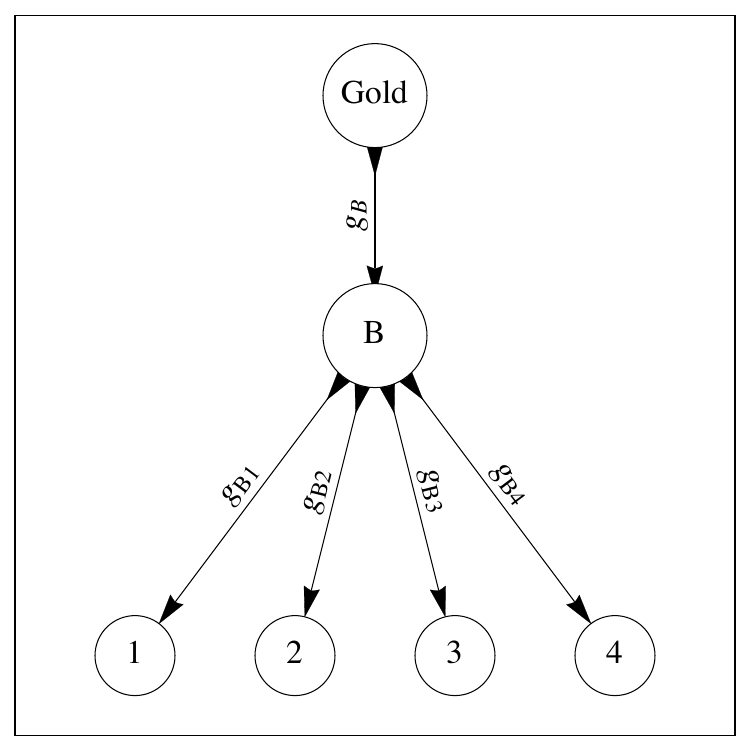}
\includegraphics[height=0.32\textwidth]{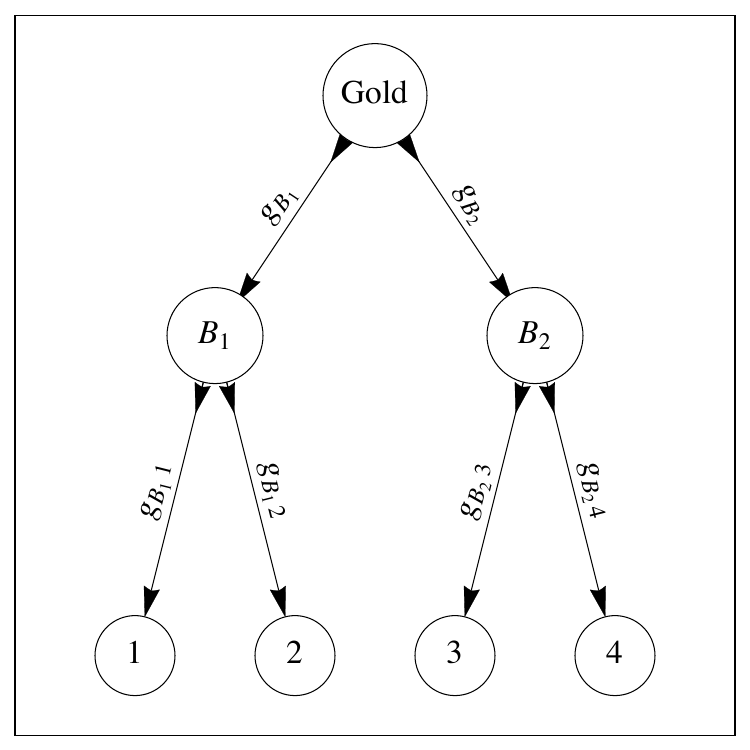}
    \caption{Left top: direct commodity ownership with
      a graph representation. Left bottom: standard representation of
      direct commodity ownerships without a graph. Middle: indirect
      ownership of a commodity through a single bank. Right: indirect
      ownership through several banks, with each bank owning directly the commodity.} 
      \label{fig4}
\end{figure}

The simplest description consists of representing the owners together
with the amounts they own. However, in order to unify the
representation of ownership with the representation of debts that we
will include later, it proves useful to depict the representation of
ownership on a graph. A graph can be thought intuitively as a set of
points, the vertices, some of which are linked by lines, the edges. The
totality of a given commodity is \CJE{represented by} one vertex and so are the various
owners, the relations of ownership being represented by an edge from the
commodity vertex toward the corresponding owner. Both
representations are depicted in the left plot of Fig.~\ref{fig4}. In
the graph representation, the edges are weighted by the
corresponding amounts owned. The resulting graph structure is thus a
weighted and oriented graph with a simple tree structure. The oriented edges can be viewed as arrows, and when departing from a vertex they represent a liability, while when incoming on a vertex they represent an asset.

A slightly more complicated structure arises if \CJE{the entirety of} the commodity is
completely stored in a bank vault. There is one
direct relation of ownership between the commodity and the bank, \CJE{with
the bank issuing} corresponding liabilities to its customers.
The members of the system either a) directly possess the UoA, or b) own it indirectly
through a bank, having a (positively credited) account in that
bank. \CJE{In scenario b}, we say that the bank issues an {\it I Owe
  You} (IOU), \CJE{as a record of} the gold placed into the vault. In
scenario a, the customers need to physically exchange the
commodity.
%, whereas in the second case they only exchange the ownership certificates (the IOUs). 
\CJE{Scenario b is expanded when these indirect ownerships are} made through several banks, as in free banking systems. We
illustrate this in Fig.~\ref{fig4}.

%DONE.

\subsection{Emergence of central banks}

\CJE{Several configurations are bound to appear once} customers of different banks start to
interact in the economic world and need to exchange to settle
payments. In a system of {\it independent banking} the banks can ignore each other. This would force every customer
  to have \CJE{an} account in each bank. Indeed, once a person accepts a
  payment from somebody having their assets in a given bank,  this
  person would be forced to open a bank account in the same bank to be
  able to transfer a liability. With {\it decentralized banking}, each customer would
  still have only one bank account in a given bank, but the banks
  themselves would possess \CJE{bank accounts with} other banks, so there would be oriented edges between the banks. In this system, a bank $A$ owes gold to its customer, either because it has gold in its vault, but also possibly because another bank $B$ owes gold
  to bank $A$. This situation can become rather complex, with a
  maximum of $N(N-1)/2$ claim-debt relations between $N$ banks. In fact, a general
  theorem that whenever there are strictly  more that $N-1$ edges
  between $N$ connected actors, there must be loops, that is paths along
  the edges of a graph which start and end at the same vertex. It can thus become optimal to simplify the networks of
  debts and claims between banks, by removing all unnecessary
  loops. The simplest solution is a system based on {\it centralized
    banking}, where we use a tree graph (a graph with no loop) of unit
  depth, with one bank at the root acting as a central bank. It is
  obvious that in that case there would be only $N-1$ edges in this
  network of $N$ banks.  As explained in the next section, any new debt added between two
 actors in a tree creates a new loop \CJE{that} can be removed to preserve an optimal tree structure.
In a last round of cooperation, the banks can decide to send their gold in the vault of the central bank, and keep a claim  on it, so as to optimize the cost of gold keeping. 

Note that in the case of independent banking and decentralized
banking, there is actually a central structure hidden \CJE{by} the
possession of gold. In decentralized banking, if all banks
settle their debts, they are back to independent banking, and they are
united through the claims they have on the commodity. So even if
there is no central bank, or no central system, the fact that only one
commodity is used is functionally \CJE{similar} to a system with a central
bank. Gold acts as the root of a tree graph. The centralized
banking system \CJE{serves} to avoid physically displacing heaps of
gold \CJE{by storing it all} at the central bank. The use of gold, with its physical exchange, is rooted in the
lack of a central organization, as it effectively replaces it.

\subsection{Payments in a tree graph}\label{SecPay}

\begin{figure}[!htb]
\includegraphics[height=0.38\textwidth]{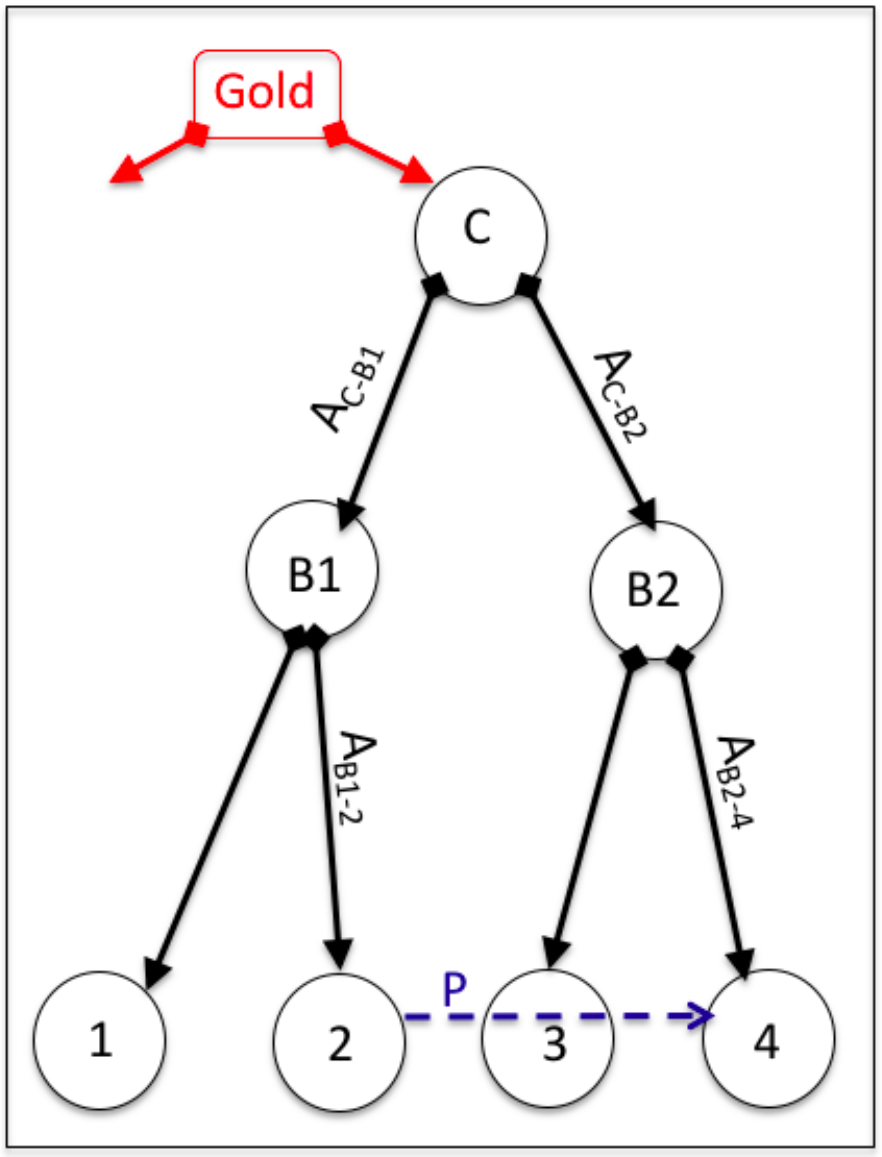}
\includegraphics[height=0.38\textwidth]{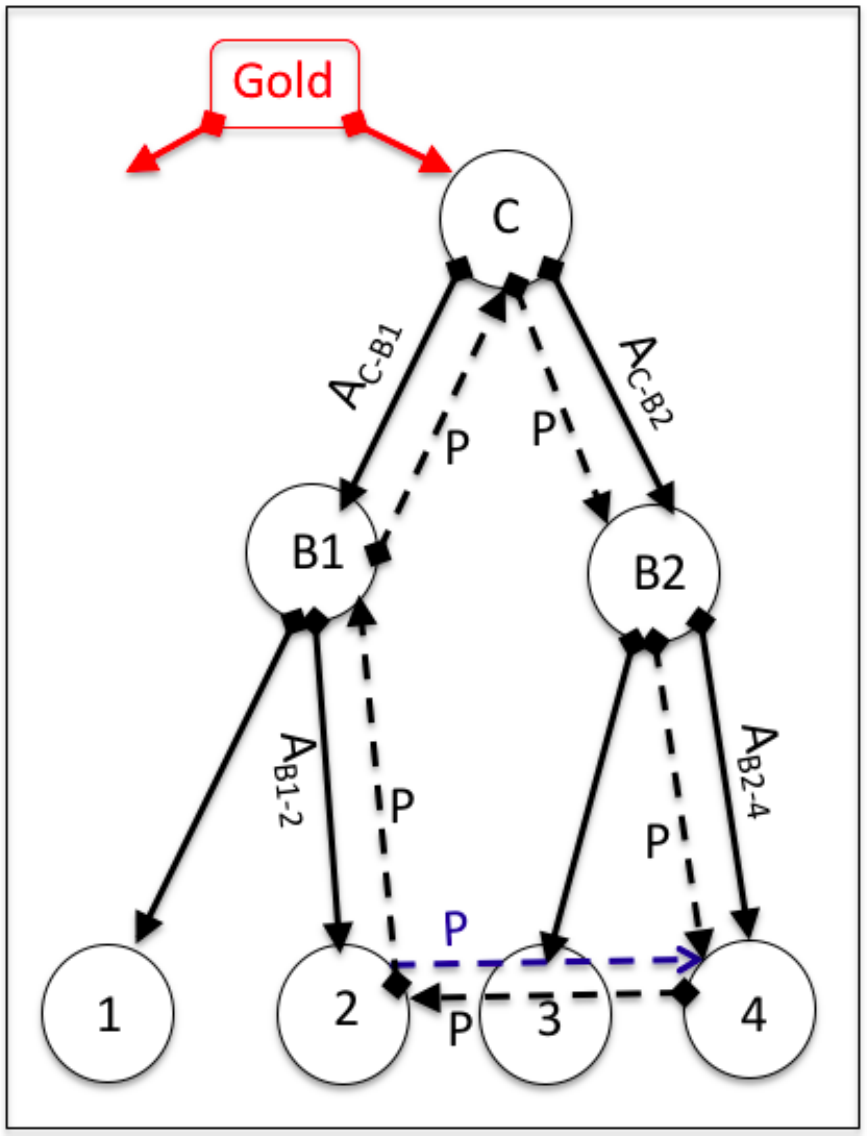}
\includegraphics[height=0.38\textwidth]{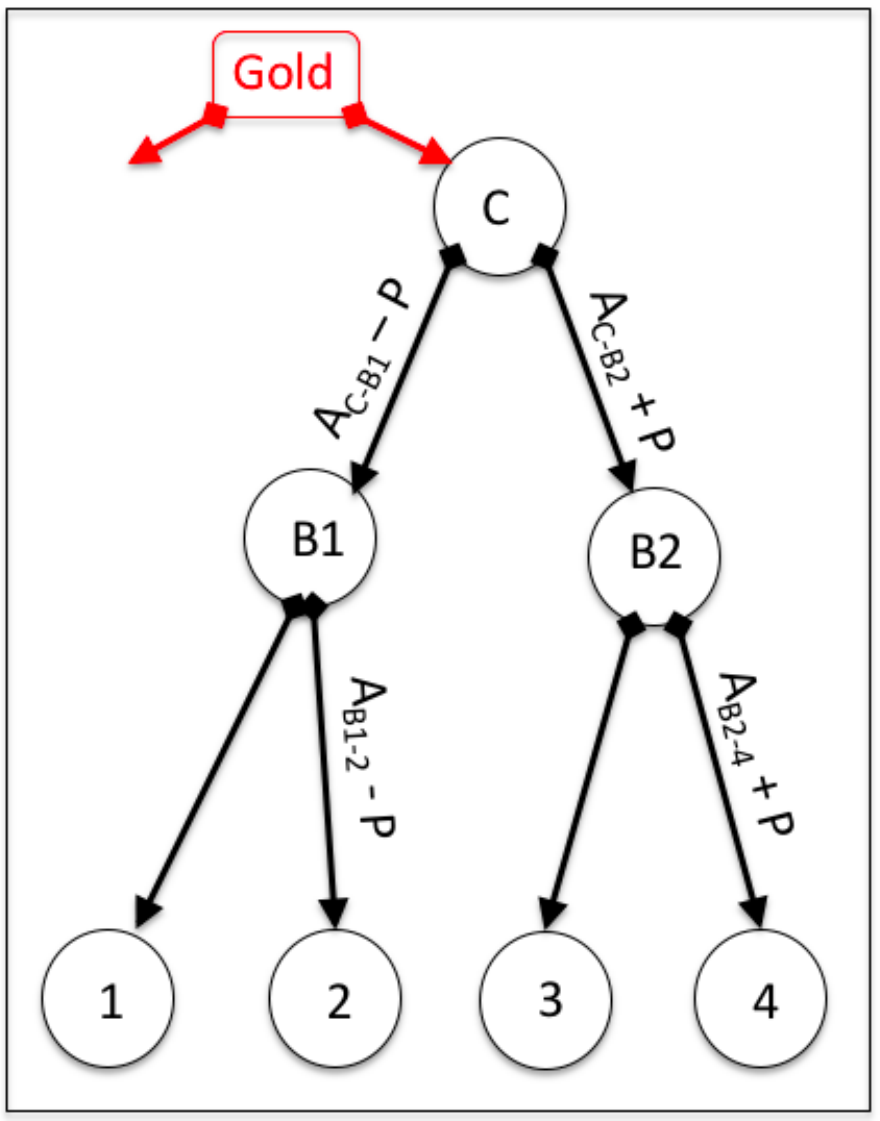}
    \caption{Description of payments via graphs. $C$ stands for the central bank, $B_1$ and $B_2$ stand for the commercial banks and
      $1,2,3,4$ are the various customers of the system.  All
      debts/claim relations are reported as weights on the edges of
      the graphs, and are expressed in the UoA of the underlying
      commodity fully stored in the central bank.  } 
      \label{fig5}
\end{figure}

If the tree-structure has been clearly identified with a central bank and all other banks directly related to
  it, then payments between customers of these banks are easily depicted. For each pending payment there is a unique loop created because the payment at this stage is the
addition of a new debt. \CJE{Recall} that a tree with $N$ vertices, has
$N-1$ edges, so when one edge is added, there is necessarily one loop.
In order to maintain the tree structure and to avoid direct claims between customers or between banks, this loop needs to be
removed.  For that purpose, when a customer $a$ pays a customer $b$,
it needs to provide its bank $A$ with the name of the bank $B$ of
customer $b$ \CJE{so that the loop can be identified}. 
%If the monetary system was previously in a tree
%form as would be the case if commercial banks do not keep claims
%between each other, this loop is unique. 
\CJE{Then,} we simply need to subtract the amount of the
payment in all \CJE{edges} of the loop as illustrated in
Fig.~\ref{fig5}. In the first panel a pending payment results in
  a new short-term  debt which creates a new loop. Then this loop is
  emphasized and in the last panel the tree structure is recovered after the loop is removed, and the new balances of
      debts-claims have been adjusted. Note that we are actually
      stacking two types of graphs. The red graph represents commodity
      ownership relations, and the black  graph is made of IOUs for that
      commodity. Note also that commercial banks can decide
  to keep direct claims as in decentralized
  banking. In that case, one would have a loop displacement rather
  than a loop removal, as the direct claim between customers would be
  transformed into a direct claim between commercial banks. It amounts
  to a partial settlement since it maintains the tree structure below each bank but
not among banks. \CJE{Note also that with this structure there are
several ways} to remove the extra loop introduced by the
payment of one customer to another one. In the remainder of this article we do not
investigate in more details the interbank structure and for simplicity
we assume that \CJE{payments} are fully settled.
At any time step, there will be payments, and by successively
incorporating them, we effectively build the graph of stocks as an
addition of payments from an initial condition. 
%Rephrased differently, the graph of payments at a given time is the
%time derivative of the graph of stocks at that given time. 
Graphs of stock are the position of the monetary system, and graphs of
flows are their variation.

%Note that the removal of the loop might involve the full tree (if the
%customers are not in the same bank for instance) but it can involve
%only a sub-tree (e.g. if the customers are in the same bank). But i
%In any case, a payment between customers requires to identify a loop that needs
%to be eliminated. For that purpose, when a customer $a$ pays a customer $b$,
%it needs to provide its bank $A$ with the name of the bank $B$ of
%customer $b$. If the monetary system was previously in a tree
%form as would be the case if commercial banks do not keep claims
%between each other, this loop is unique. However if banks perform only
%partial settlements of payments and thus always keep claims among each
%other, there can be several possibilities to remove the extra loop introduced by the
%payment of one customer to another one.

%DONE

\subsection{Representing debts}\label{SecDebts}

Debts from customers of the system toward a bank can be incorporated in
this framework, as it is a liability for the member and an asset for
the bank. They are all based on the same principle: to lend an amount and to ask for a higher repayment at maturity. The difference between the initial amount lent $S$, and the repaid
amount can be converted into an interest rate $i$ for a given unit of
time, using typical \CJE{compound interest} for the duration of the maturity $D$. 
%It can sometimes be more convenient to work with a log-interest rate defined by $i \equiv \log (1+I)$. The amount to be repaid
%is then $S \times {\rm e}^{D\,i}$. Of course if $I$ and $D$ are small,
%  the amount to be repaid is approximately $S (1+ D i ) \simeq S (1+ D I )$.
Any complicated loan structure can always be considered as a set of
 independent loans \CJE{following this same simple structure}. 
%(e.g. for a mortgage where the repayment has to be made every month in say $240$ month, it can be considered as the addition
%of $240$ independent loans), all with the same interest rate, but with
%different maturities, which have been arranged such that the
%repayments are always kept constant. For the sake of understanding the nature of
%debts and interests, we will thus always ignore complex schemes like
%this and consider a single maturity for an amount borrowed. 
In the graph representation, we indicate a debt by $(S,i,D)$
where $S$ is the current amount, $i$ the interest rate and $D$ its
maturity. 
%Equivalently, one could decide to represent the current amount owed, the maturity, and the higher value
%at maturity.  
After every unit of time, $(S,i,D)$ is replaced by $(S\times(1+i),i,D-1)$, and when the maturity becomes null, it is
transformed into an immediate payment of amount $S\times(1+i)^D$ whose
loop needs to be removed exactly as described in
~\S~\ref{SecPay}. And just like for a payment, it comes as a reduction of the liability from the bank toward the member that the latter must have
  earned through previous payments made to him. Because the nature of
  each debt is different due to variations in interest rates and
  maturities, this means that we are now considering several types of
  assets and liabilities \CJE{within the same graph}. By representing several types of debts on
  the same financial graph, we are actually stacking several weighted
  graphs on the same set of vertices.  The representation of debts is used in \S~\ref{SecLoanDeposit} where we
detail credit creation. 

\subsection{Money and net money}\label{SecNetMoney}

%This conservation law calls for a clear difference between net money and
%money, which are respectively the analogues of total charge and
%charge displacement.

The IOUs of the bank are \CJE{merely a} promise to convert them into IOUs of the
central bank (and if there is convertibility, these IUO can be
converted into gold). These financial assets are equivalent to interest-less debts. By introducing the
possibility to have debts, we see that we will have a mixed system \CJE{with} two categories
of IOUs. In the first category, we have the IOUs of the
central bank, and the IOUs of commercial banks which are promises to
deliver IOUs of the central banks, and in the second category we have the IOUs of customers who have
borrowed and \CJE{who} promise \CJE{to repay their debt} in the future using IOUs of the first category. Throughout this paper we will make a
distinction between {\it money} and {\it net money} when referring to
these IOUs:
\begin{itemize}
\item money held by a customer of the system consists in the financial assets, and it
  is the reflection of IOUs of commercial banks and of the central bank
  toward the customer;
\item net money held by a customer consists in money from which we
  subtract any liabilities, hence it \CJE{consists of} financial
  assets minus the customer's financial liabilities, or net assets. 
%If there is gold convertibility and no Treasury, net money is equal to the amount of gold held in the central bank.
\end{itemize}
These definitions are expanded to a group of members
of the system by direct summation. 
%All amounts, be it for money or for net money, are expressed in the
%same UoA, e.g. the national UoA set by the central bank. 
As we shall detail in \S~\ref{SecLoanDeposit}, any new loan modifies
the total money, but conserves the total net money. \CJE{Therefore
attention must be paid to using this precise language.}
\ifmoneyagg
This distinction between
money and net money is also related to the various definitions of monetary
aggregates that we discuss in \S~\ref{SecMi}.
\else
\fi

%DONE

\subsection{Consolidations and sub-graphs}\label{SecConsolidations}

\begin{figure}[!htb]
%\begin{center}
\includegraphics[width=0.42\textwidth]{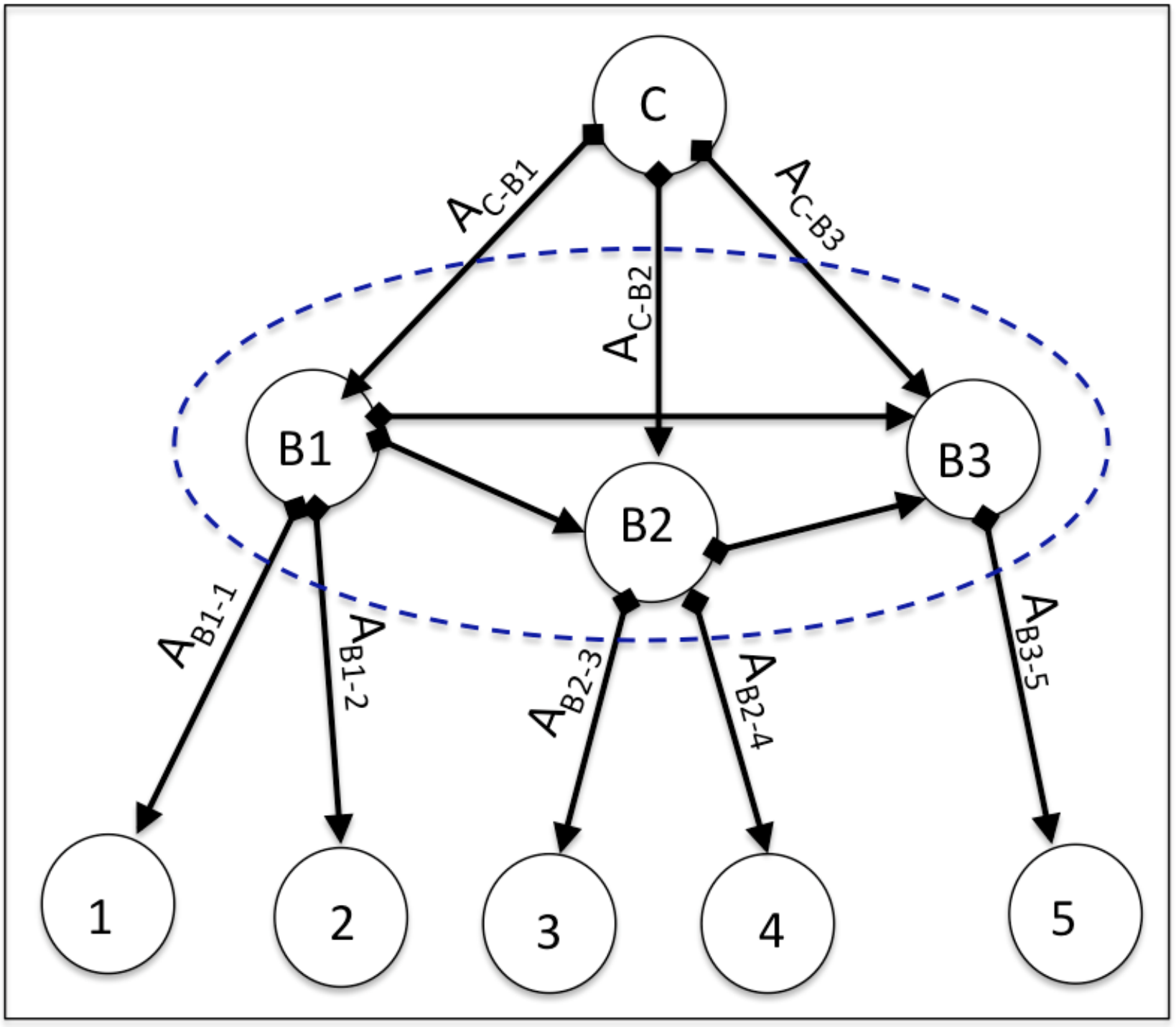}
\includegraphics[width=0.42\textwidth]{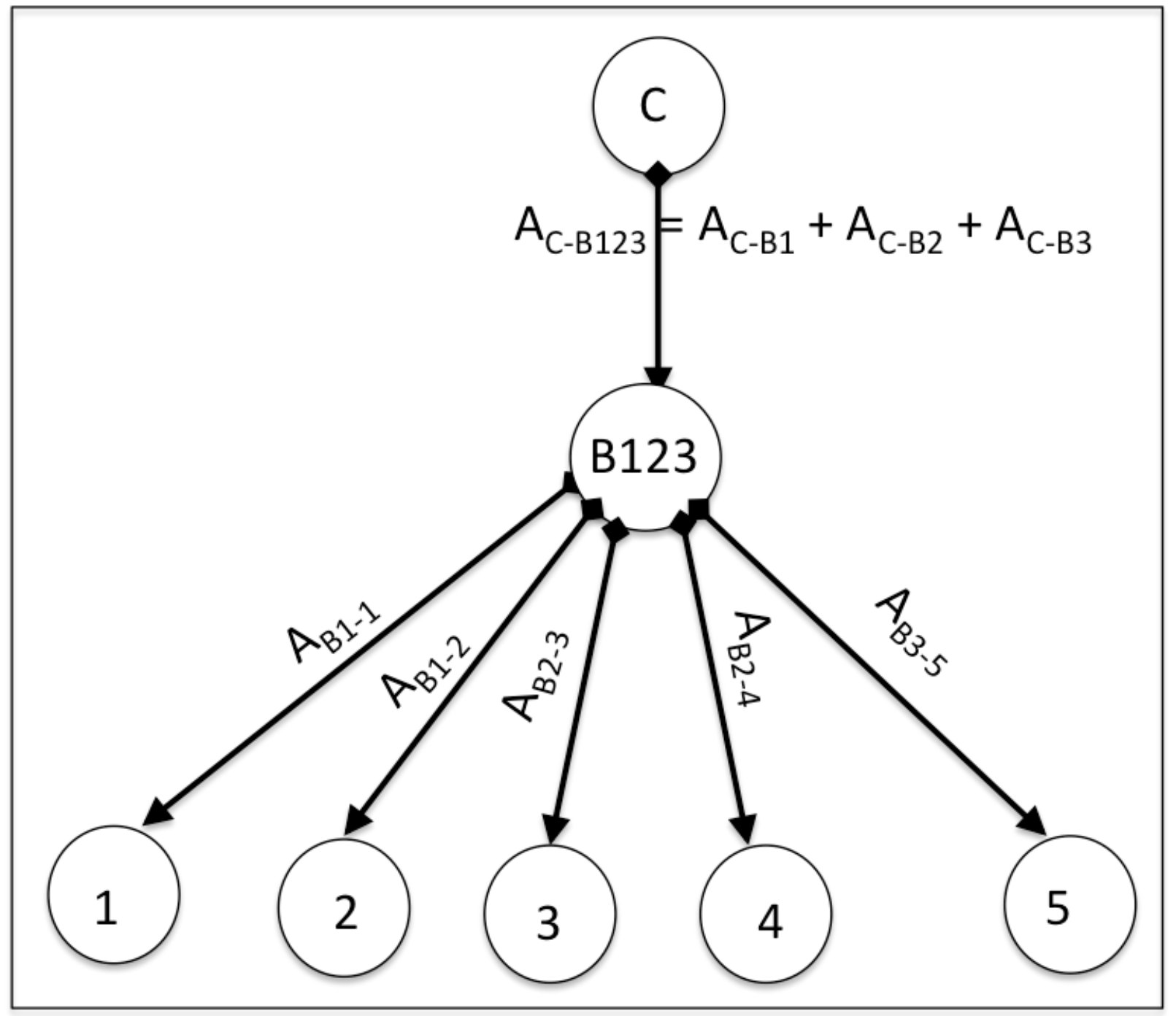}
%\end{center}
\caption{Left: the closed curve identifies the commercial banking sector. Right:
  consolidation of the commercial banking sector.}
   \label{fig7}
\end{figure}

Let us draw a closed line in the monetary graph to define a sub-region. We can replace the subregion by a single
vertex whose assets and liabilities are depicted by the incoming and outgoing lines of the
region. In practice, \CJE{this} means that if we do not need to describe with precision the
internal financial structure inside the region, we can ignore it and simplify
the graph by \CJE{using a} simple vertex. In accounting, this is called a {\it consolidation} and helps in reading the structure of graphs. One can for instance decide to consolidate the
financial sector, and thus to reduce all banks to just one bank like
in Fig.~\ref{fig7}. Through consolidation, all financial relations
which start and end inside the consolidated region do not matter for
the exterior, so when the banking sector is considered
from a consolidated point of view, interbank relation do not matter at
all. %Actually nearly all interesting graphs are
%consolidated, so as to show a specific feature of the system.

Alternatively, instead of erasing all the details of the
  sub-region, we can focus on it by eliminating the
  details of the outside region. This amounts to considering a
  sub-graph, with externals sources and sinks. For instance, in the
  left plot of Fig.~\ref{fig7}, we could focus on the banking sector
  which sits inside the closed curve. This is exactly what is done by \citet{SheldonMaurer,UpperWorms,Upper2011,Gai,Wells2004} where the emphasis is on the stability of the banking sector itself.

\subsection{Relating balance sheets and financial graphs}\label{SecMatrix}

We have \CJE{provided} the basic tools to represent a financial system with
graphs. Since a financial system can also be represented using balance
sheets, \CJE{it is worth discussing why} these two approaches are related
and why graphs are more convenient. 
\begin{itemize}
\item {\it Equivalence with balance sheets.} A weighted graph is equivalent to
an antisymmetric matrix $W$, whose entries correspond to the weights of the graph. For instance, if we have
$W_{ij}=-W_{ji}>0$, then this value corresponds to the weight of the directed edge from the vertex $i$
toward the vertex $j$. Balance sheet representations are then just a
special way to visualize such antisymmetric matrices. \CJE{They amount to} detailing the matrix column by column, separating the positive and
negative entries. For a chosen vertex $j$, the entries of the column
are the $W_{ij}$ where $i$ runs over all vertices. If $W_{ij}>0$, then
the financial relation is an asset as it corresponds to an IOU from
vertex $i$ to vertex $j$, and if $W_{ij}<0$ it is a liability since it corresponds to an IOU from vertex
$j$ to vertex $i$. Said differently, for a given vertex the outgoing
edges (resp. the ingoing edges) are the liabilities of that vertex
balance sheet (resp. the assets of that vertex balance sheet). For
each graph representation, the balance sheet picture can be inferred by the examination of the edges attached to each
vertex. Conversely the graph and its corresponding matrix can be
reconstructed from the list of all balance sheets. We must note that
if we have several different types of financial relationships, then we have several types of antisymmetric matrices and thus several types of weighted graphs and
several types of entries in the balance sheets. For instance in Fig.~\ref{fig5}, we have one graph for the ownership of
gold \CJE{that} is partially represented, and one graph for IOUs on
gold.  \CJE{In the same way that} matrices can be added, graphs can be stacked, and balance sheets
can be combined so as to lead to a global representation of the
financial relations.
\item {\it Advantages of graphs.}
It is simpler to visualize the global properties of a financial system from a graph
representation than with just a list of balance sheets. Indeed with
balance sheets the related entries $W_{ij}$ and $W_{ji}=-W_{ij}$
appear twice, since they appear in the balance sheet of vertices $i$
and $j$. For instance if $W_{ij}>0$ it appears in the liabilities for
vertex $i$ and in the assets for vertex $j$. The reader must \CJE{mentally reconnect}
all these pairs to obtain a clear \CJE{understanding} of all financial \CJE{relationships}. With a graph, the entries $W_{ij}$ and
$W_{ji}$ appear naturally as a single weighted edge connecting vertices $i$ and
$j$. \CJE{Financial relationships} are thus directly observable without any
effort and consolidations are immediate as described in
\S~\ref{SecConsolidations}. The balance sheets representation is the
collection of the individual point of views on a monetary arrangement,
whereas the graph representation is the global point of view which
allows to grasp the emerging features of the system. If a balance
sheet is enough for accountants needing only their local situation
with respect to the rest of the monetary arrangement, the graph is
necessary to visualize and thus understand the global organization.
\end{itemize}
%TODO HERE

\section{Private banking and endogenous money creation}\label{SecEndo}

Having all the tools to \CJE{graphically represent} financial stocks, we
are now ready to investigate the general organisation of monetary
systems. In this part, we focus on the specificities of the banking sector and
postpone the role of the Treasury to \S~\ref{SecState}.

\subsection{Net worth of banks}\label{SecNW}

In this section, we define the net worth of banks so as to be able to investigate in detail the mechanism of
money creation through credit creation in the next section. 
An immediate issue when allowing for debts, is the possibility that
the borrower might not be able to repay \CJE{the loan when it is due}. Indeed, the borrower thinks that he will be able to repay in the future
because he is going to exchange his work against money, \CJE{which} in turn
he will use to reimburse. Once the debt is turned into
an immediate payment at maturity as explained in \S~\ref{SecDebts},
the borrower must have sufficient assets \CJE{in order to be able to repay
the loan.} 
%There are then obvious reasons why the borrower might not be able to work to find the money needed. 
%%This ranges from losing his job, having an accident, or simply being
%%dead, among many other reasons. 

From a graphical point of view this would be equivalent to a
unilateral removal of the debt from a customer to its bank. However,
if the bank had initially as many liabilities as assets, this is no
more the case after the debt removal, so in general \CJE{one} cannot assume
that liabilities and assets are balanced. We thus introduce the
concept of {\it net worth} of a vertex, that is a bank, which is the difference between assets and
liabilities of that vertex. 

So far the bank was just a node or a vertex in the graph structure,
but behind a bank there is a banker\footnote{More
  realistically the physical group of persons made of
  stockholders. But throughout this paper we symbolically describe
  these as {\it the banker}.}. If a borrower defaults, the banker
should pay the corresponding debt for him and this takes the form of a
corresponding decreasing of the bank net worth. Conversely, when the borrower reimburses but also
pays interest, it does so to the banker and this means that the bank
net worth is continuously increased by the interest, and
  this is also another reason why the introduction of a net worth is
  necessary. To summarize, the risk of defaulting together with the benefit of charging
interest, are affecting in opposite directions the bank net worth. 

The notation of consolidation needs to be extended to take into account
 the net worth of banks or companies. If we draw a closed region that we want to
 consolidate, we just need to add all the values of the net worths
 which are inside, and they will add up to the equivalent net worth.

%DONE
\subsection{Loans, deposits and banking}\label{SecLoanDeposit}

%An everlasting debate is about knowing whether or not deposits allow
%customers to borrow, or if it is because customer borrow that there
%are deposits. This question is reminiscent of the debate about who was first there between the egg and the hen. We will tell two stories, one where it is
%the egg in this section, and one where it is the hen in the next
%section, which will then allow us to conclude that these stories are
%harmless as long as they help thinking the possible underlying
%mechanisms, but they are not fundamental at all.

\begin{figure}[!htb]
\includegraphics[height=0.315\textwidth]{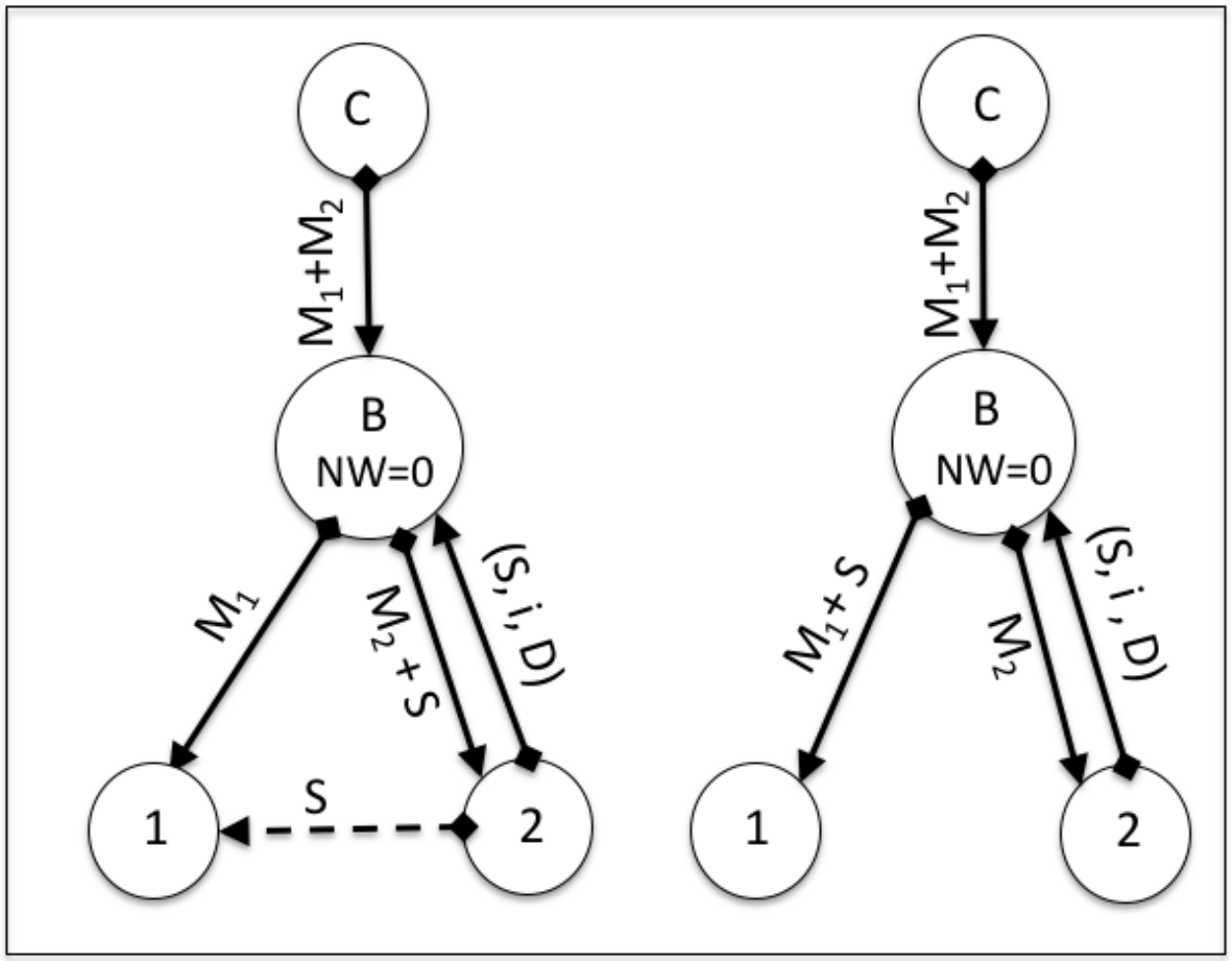}
\includegraphics[height=0.315\textwidth]{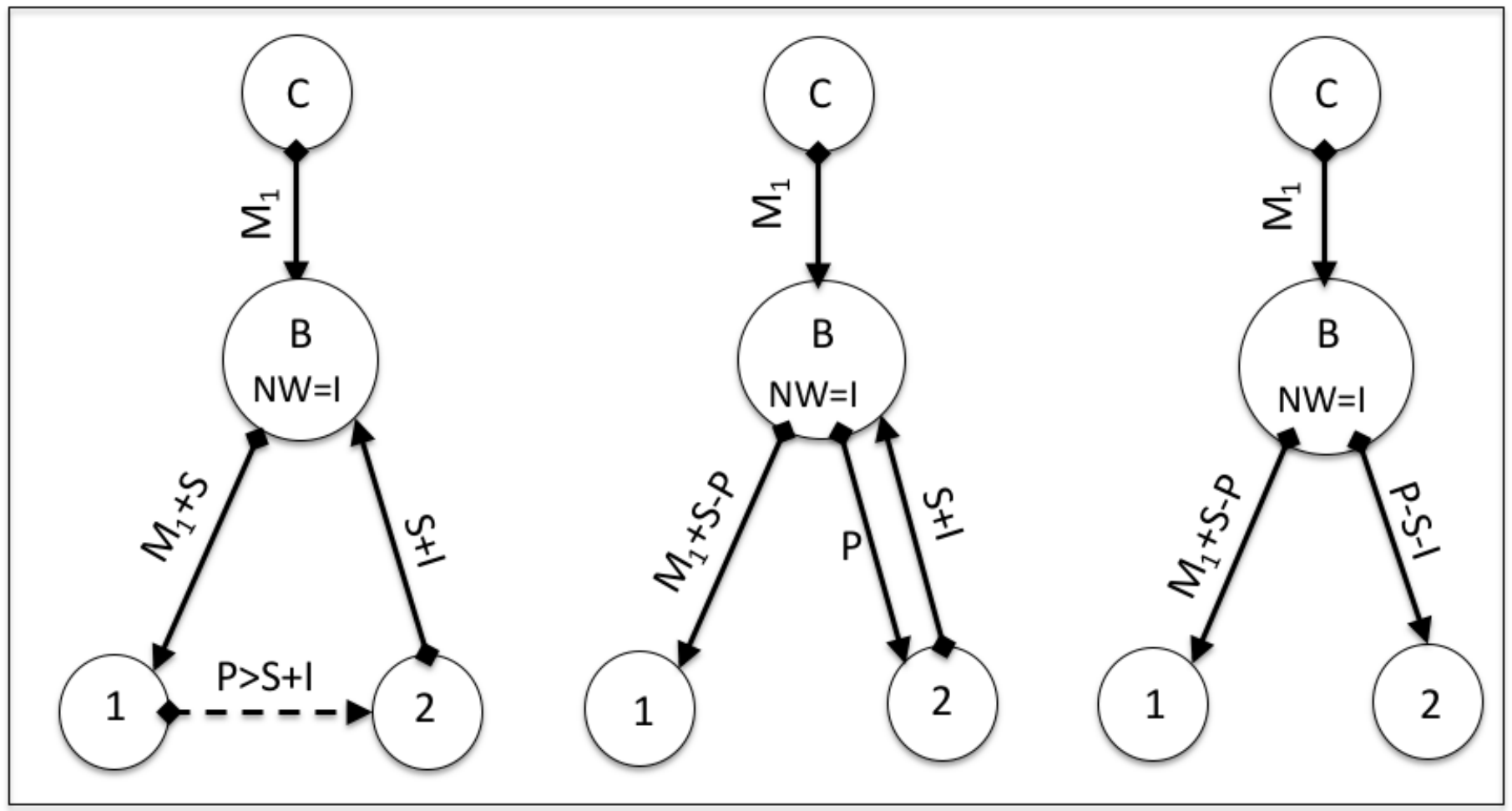}
\caption{{\it Left}: Creation of money after a loan is granted and spent.
{\it Right}: Loan repayment and destruction of money. }
\label{fig8}
\end{figure}

In this section, we detail the process of credit creation, and
illustrate the well known fact that it is at the origin of money, as
bank loans create deposits~\citep{Tobin}. This is illustrated in the left plot of
Fig.~\ref{fig8}.  In the first graph of this plot, the customer \CJE{initially takes out} a loan for an amount $S$,
 with interest rate $i$, and maturity $D$ which is represented on the
 edge going from customer $2$ to bank $B$. \CJE{As a result, customer $2$'s
 liability to bank $B$ has} increased from $M_2$ to
 $M_2+S$. The bank $B$ also has liabilities toward customer $1$ for an
 amount $M_1$. This is reflected \CJE{in} the liabilities from the central
 bank $C$ to the bank $B$ which sum up to $M_1+M_2$. The payment that customer $2$ intends to make to customer
  $1$ is superimposed as a dashed arrow.  In the second graph, after
  settling the payment, customer $2$ is left with an extra liability
  and customer $1$'s assets have increased. We observe that the sum of
  all customers' assets and liabilities owed to the bank is
  unchanged. Indeed, if we were to consolidate the customers into only
  one customer, then the total net assets, that is the total debt of the banking sector toward the
consolidated customer is unchanged. But since we have decided to call
money the financial assets of customers, and not the
net assets, we can say that there has been money creation as a result
of the act of borrowing and spending. 

The reverse process occurs when the loan is to be repaid after $D$ units of time have
elapsed. This is illustrated in the right plot of Fig.~\ref{fig8} where we assume for
simplicity that the customer has no money ($M_2=0$). 
In the first graph the borrower needs to work so as to receive a
sufficient payment which is settled in the second graph. Since
the borrower owes the principal plus interest, the net worth of the
banker has increased by an amount $I$ equal to the \CJE{expected
interest}. In the third graph, the borrower repays the loan and credit
money is destroyed. The net worth of the bank has been increased by
the interest $I$ which \CJE{has} been subtracted from the total of the
customers' assets. The liability from the central bank to the
commercial bank is unchanged. Eventually the bank can redistribute
totally or partially its increased net worth among its employees and
shareholders in the form of salaries and dividends.
This dynamic creation and destruction of money is called the {\it endogenous } nature of money.

It is often not intuitive to understand that banks can create
  money, since one would expect that they would create arbitrary
  amounts of money for themselves. However what banks can create
  endogenously is {\it money}, not {\it net
  money}. That is if they create money and give it to their bankers and
  shareholder it comes as a decrease of their net worth, as is evident
  if they then spend it to pay customers of other banks. It is equivalent to an advance on future salaries and dividends, and banks can only do so with the restriction that the net worth
remains positive.

%There is no net creation of assets because if we have increased the assets, we
%have also increased the liabilities. 

%Considering money and not just net money, that is considering only the
%financial assets of customers, amounts to looking at the displacement
%of money inside the customers sector.
% There is a point where there is negative money, that is liabilities of the customers. The loan has
%resulted in the creation of money, and

%To conclude, in this story, it is the loan which has made the
%deposit. It has created money but since it was just there to polarize into equal amount of
%assets and liabilities it has not created any net money.  We can also
%see that some simple fundamental questions arise. What if all
%customers with positive assets ask to convert these assets into the
%central bank IOUs? What if the borrower pays somebody in a different
%bank and not in the same bank? 

%DONE

\subsection{Interbank payments and financial institutions}\label{SecIntBank}

\begin{figure}[!htb]
\includegraphics[height=0.335\textwidth]{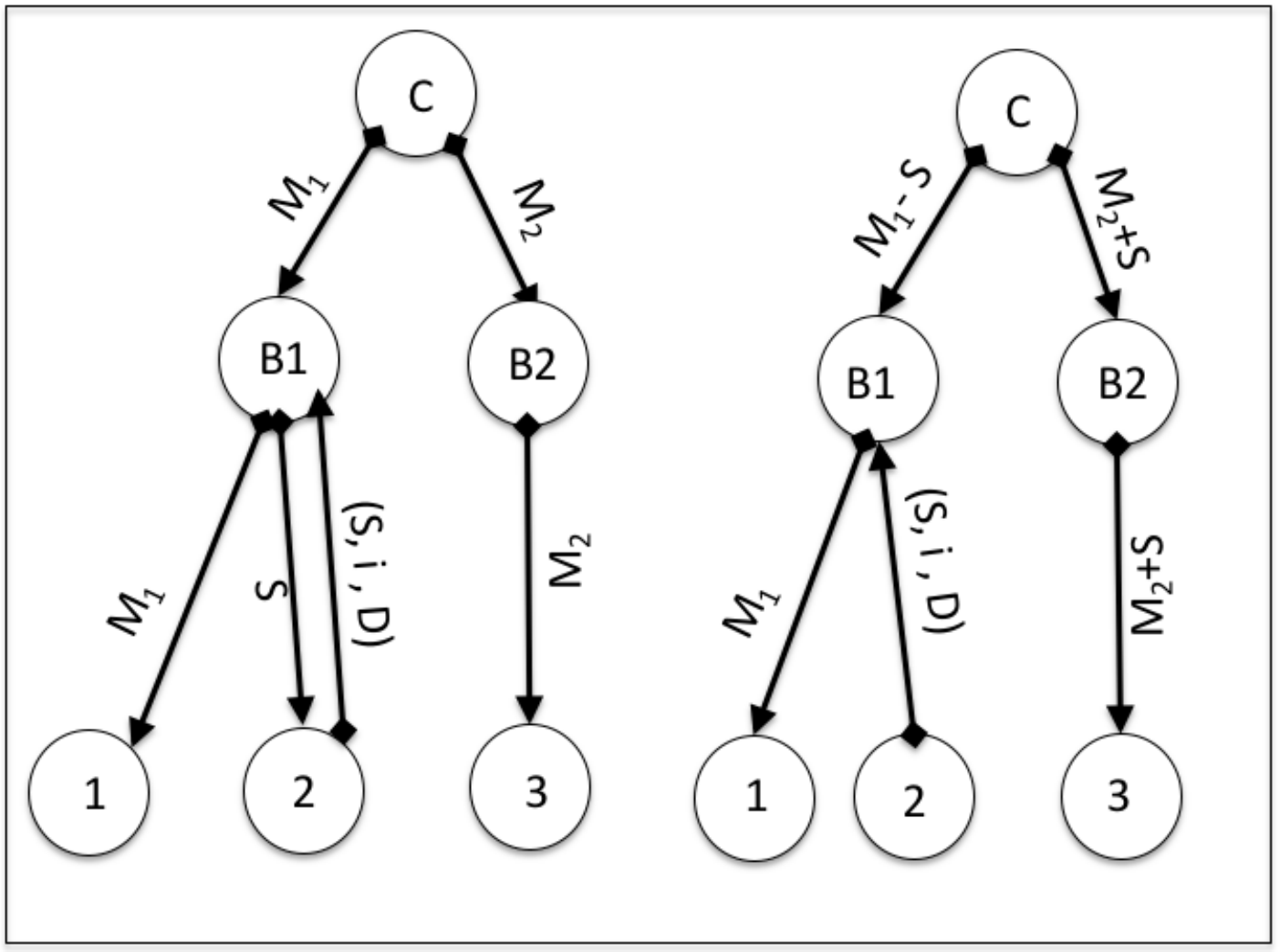}
\includegraphics[height=0.335\textwidth]{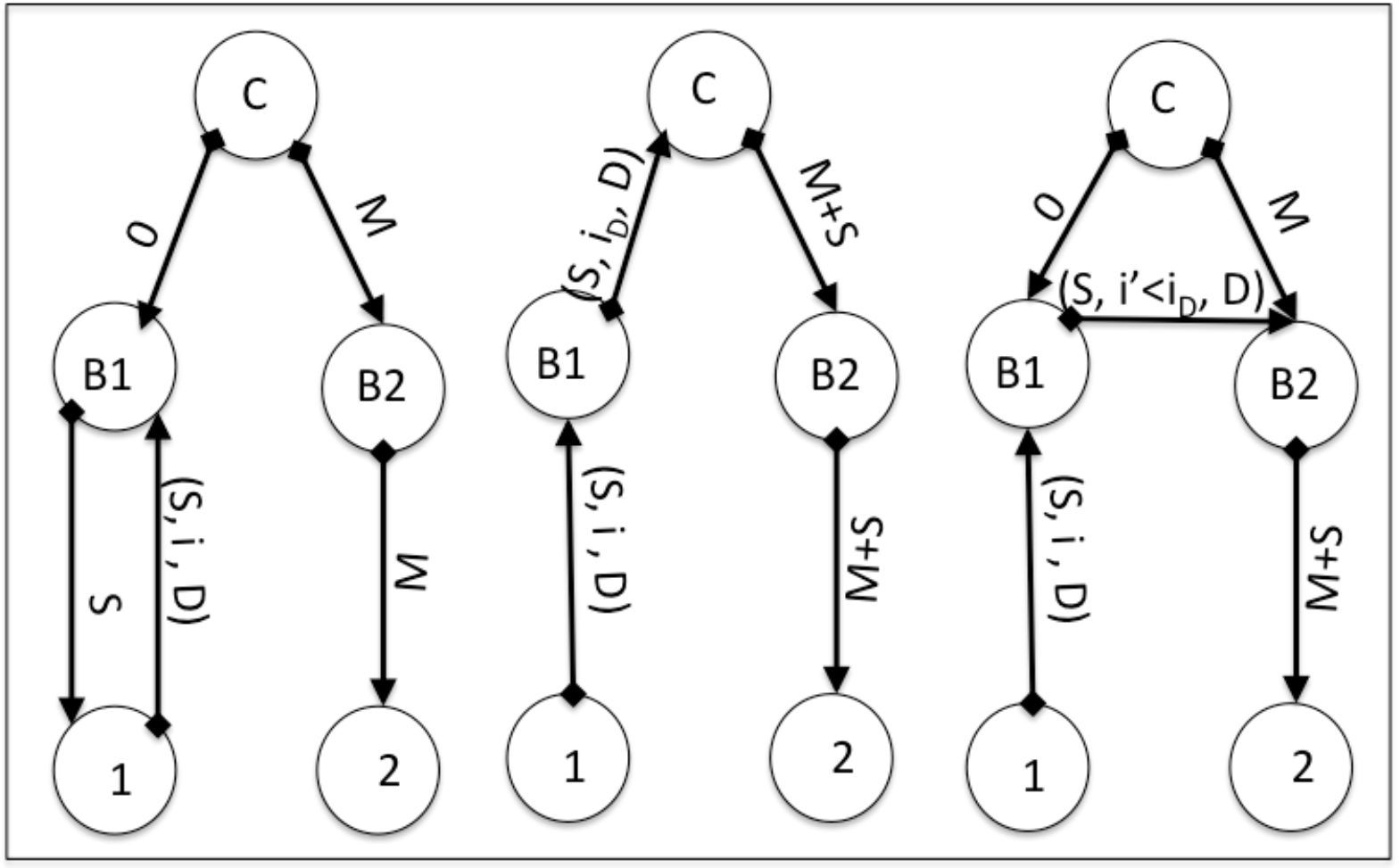}
\caption{A loan is granted and spent to pay a customer of a different
  bank. In the left plot, the bank which granted the loan has enough central
  bank reserves, and in the right plot it does not have central bank
  reserves at all, so it must either borrow to the central bank or in
  the interbank market.}
\label{fig11}
\end{figure}

In the previous section, we considered customers who could pay each other
easily because they have accounts at the same bank. If the borrower spends
by paying someone with an account at another bank, this creates a loop in the
banking tree structure, which is removed when the payment
is actually settled. This is straightforward if the borrower's bank
has enough central money as illustrated in the left plot of Fig.~\ref{fig11}. In the first graph the bank grants the loan and
creates the corresponding deposit. In the second graph we illustrate
the result of the payment to a customer in a different bank. This way of lending is typical of financial institutions since in order to have
sufficient central money they must have received earlier deposits from customers.

However if the borrower's bank has no central bank money because
it never received deposits from customers, this
proceeds differently. In the right plot of Fig.~\ref{fig11} we
assume that the bank \CJE{making} the loan has no central bank money when
it grants the loan in the left graph, but it makes sure to ask for an interest rate higher than
the discount rate. This leads to two possibilities once the payment to a customer of a different bank is
settled. The first possibility is the borrower's bank is indebted
\CJE{to} the central bank, which is illustrated in the middle
graph. The central bank will charge interest at the discount rate for
this and acts with banks as a bank acts with its customers. The second
possibility is that since the second bank now has an excess of central bank
money bearing only the deposit rate, it might prefer to lend
it directly to the indebted bank at a higher rate but \CJE{that} is
still less than the discount rate, such that both banks benefit from
this arrangement. We have not written the net worths of bankers for
simplicity, assuming they vanish, but they will positively evolve \CJE{due
to an accumulation of} interest payments.
%We recover a decentralized banking system in which they all have mutual debts as there
%is no tree structure in the commercial banking sector.}

Importantly, since the central bank guarantees the payment of interest
on the deposits (possibly zero) of commercial banks at the central bank, but also guarantees that the commercial banks can borrow from the
central bank at the discount rate which is of course higher than the
rate paid on deposits, then no commercial bank would loan
its excess of central bank money at a rate lower than the deposit
rate, and would borrow at a rate larger than the discount rate. The central bank thus
defines a corridor of rates in which the commercial banks negotiate
interbank loans. The interbank rate evolves on a daily basis and the
central bank tries to influence it with its monetary policy (see ~\S~\ref{SecMonetary}).
%The discount rate is just meant to safeguard against a dysfunctional interbank market and guarantee that commercial banks can always borrow to settle interbank payments.

If the bank of the borrower had enough central bank money (due to
previously existing depositors) to settle the interbank payment, it acts like a {\it financial
  institution}, as we could think of it as lending the previous deposits of customers. But if the
depositing customers decide to withdraw their assets before the loans
made are repaid, then the balance sheet of this institution looks
again like a usual bank. The difference between a financial
institution and a bank is \citep[p10]{Tobin} {\it ’of degree not of a kind’}.
Finally, we note that the power lies in the institution \CJE{that} allows to be indebted. When a bank goes into debt, the central bank
sets the conditions of the loan. In a tree structure, we have a hierarchy with the upper vertices holding a form of power on those situated
below.

%TODO Say that people can borrow directly from each other in a wide
%sense. This are private bonds (company bonds). ABS is essentially
%banks transfering debts to customers, such that customers start
%acting as banks without even noticing it. Their net worth is at risk
%as they behave as banks. Everybody is a bank, but some are not aware
%of it which is dangerous

%DONE 

\subsection{Bank runs and the nature of money}\label{SecBankRun}

Let us show why once credit has been included in a
monetary system, it becomes incompatible with gold convertibility. 
In this scenario, the central bank IOUs are the promise to
deliver gold. In a monetary arrangement, the sum of net worths of all
vertices, that is banks, and leaves, that is customers, reflects all the IOUs issued by the
central bank. This is because if we consolidate all banks and
customers into a single member, then its assets can only emanate
from the central bank given that the consolidation erases the internal
structure. If there is no debt at all, and customers have only assets,
these assets are exactly reflected by the IOUs of the central bank
and everybody can ask at the same time to convert their wealth into
gold. The entire tree would vanish, and we would be back to a
situation where everybody has \CJE{physical shares} of the commodity. 

\begin{figure}[!htb]
\includegraphics[height=0.36\textwidth]{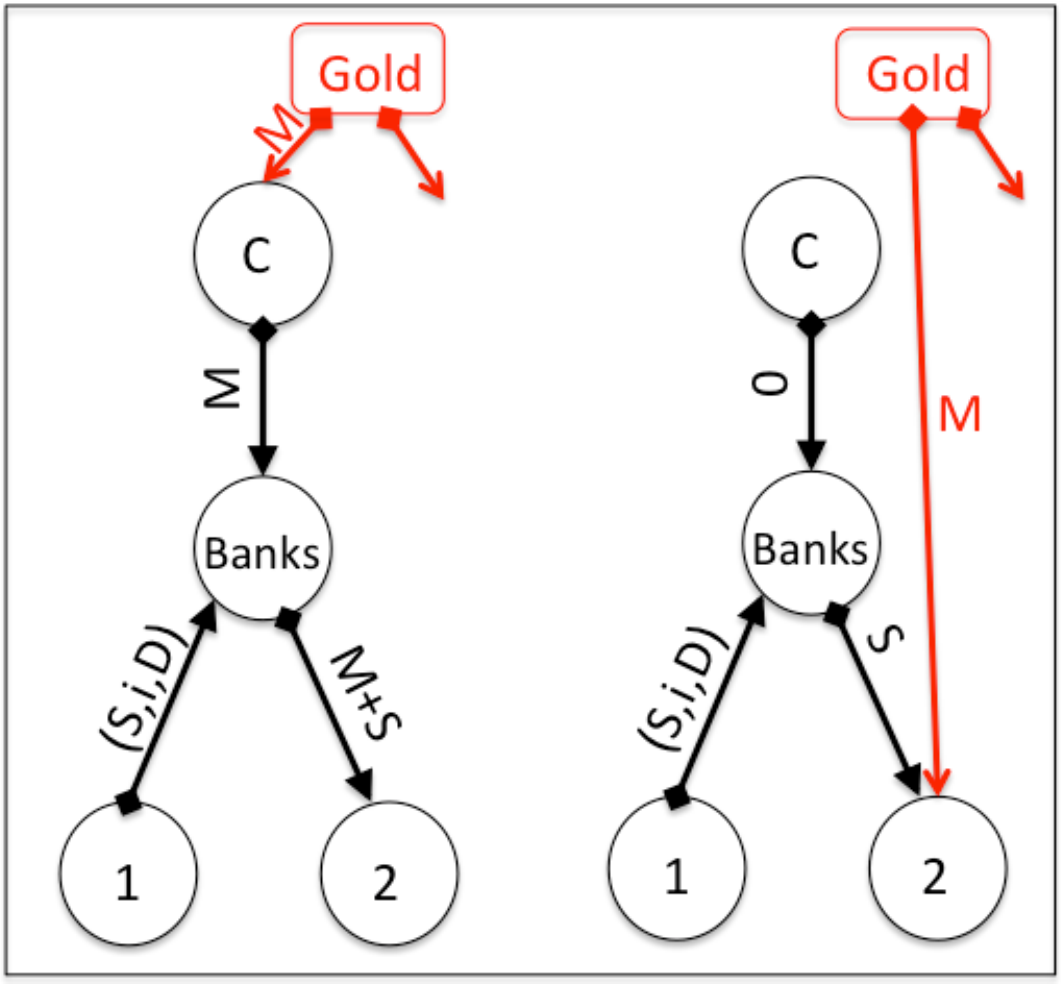}
\includegraphics[height=0.36\textwidth]{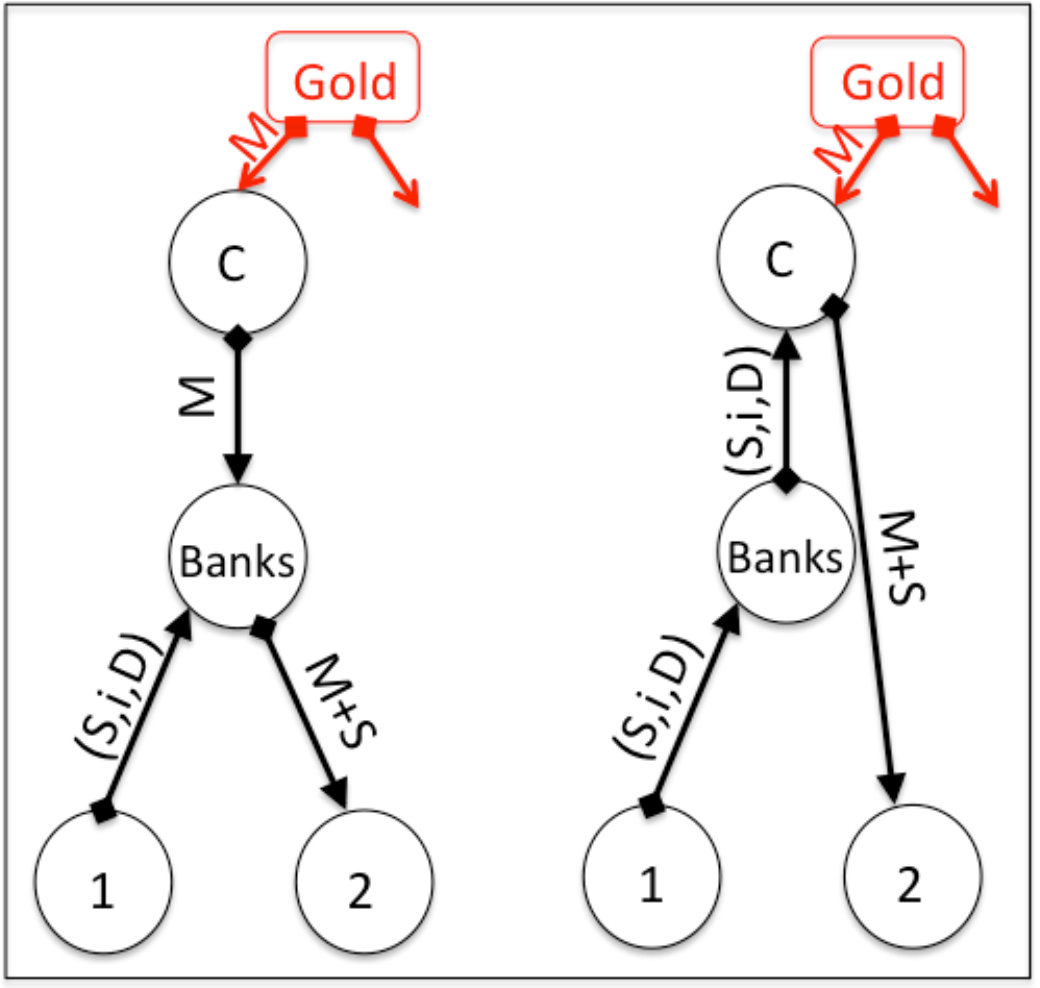}
\caption{{\it Left}: Customers ask to convert their assets into gold
  until the central bank runs out of gold. {\it Right}: if the central bank stops gold
convertibility, then customers can only convert their assets
(commercial bank IOUs) into anonymous paper money.
}
 \label{fig10bis}
\end{figure}

However, if we allow for debt among the customers then we must be careful with
the fact that money (positive assets) is not net money. If we allow all
positive assets to be converted to gold, then we have more claims on
gold than gold itself, and the system fails. \CJE{How is failure possible} if we
have been so careful to ensure double entry bookkeeping at every stage of
our monetary system? It is extremely simple and reflects the true
nature of money with respect to net money. Money is now \CJE{both}
claims on the gold of the central bank, and claims on the debts of
the borrower. It has shifted from its {\it commodity nature} to its
{\it credit nature}. The difference lies in the fact that the debts of
borrower are not available immediately since borrowers have to work to
be able to repay them. This is illustrated in Fig.~\ref{fig10bis}. In
the left plot the central bank holds $M$ units of gold which are
reflected  into the central bank IOUs toward the commercial banks. The
total assets of customers are partially reflected by the gold held at
the central bank, and partially by the debts of the borrowers. The orphan edge emerging from the total gold commodity
stands for all the gold held outside, e.g. in foreign countries, or
held privately, and even the gold which is lost or not yet
discovered.  If customers ask to convert their assets into gold,
they cannot convert all of it. At some point, the central bank needs to stop the
convertibility. In the right plot,  if the central bank stops gold
convertibility, then customers can only convert their assets into
central bank IOUs, usually in the form of anonymous paper money.

We interpret this saying that money is partially a right to get gold, and partially a right to ask for the
borrowers to work by purchasing their workforce, \CJE{even though this
right does not exist in legal terms}. Indeed when a
customer with positive assets pays a borrower in exchange of work, the reduction of the loop reduces the assets of the payer,
and reduces the liabilities of the borrower. There is net
money conservation, but there is money destruction.

It is because of this incompatibility between money and net money, the fact that
the first one is only partially a claim on the assets of the central
bank, but also partially a claim on the future workforce of borrowers,
that convertibility must be abandoned. By allowing customers
to go into debt, but by incorporating these debts in the monetary
structure, we have shifted the nature of the IOUs in an irreversible
manner \CJE{that} calls for the end of convertibility. Conversely, if no
bank \CJE{agreed} to generate money endogenously, and if loans where made by
customers to customers in the form of bonds that would not be
encapsulated into the tree structure, then the IOUs of the banks would
still be claims on the central bank gold reserves. One would have two types of
currencies. One would be the IOUs of banks which are eventually
redeemable in central bank assets (gold), and then the IOUs of
individuals that we could also exchange to settle payments. In some cases we would exchange claims on gold, and in other cases claims
on future workforce, but nothing guarantees that they would be traded at par. By incorporating the loans, or at least a major
part of the loans in the tree structure, we have obtained
a much more liquid system, at the price of abandoning
convertibility. As we explain in \S~\ref{SecPeg}, the same process
happens with a currency pegged to an external one, since the external
currency acts like gold. Allowing \CJE{the conversion of money instead of} just net
money in the external currency, contains the same internal contradiction leading to an
unavoidable breakdown of the currency peg. In the remainder of this
article, we assume that gold convertibility by the central bank is no
more possible, as is the case for the current monetary systems.
Finally, note that the hierarchical structure has already been
  discussed in earlier literature (see e.g. \citet{Mehrling}) where it has been
  understood that it emerges from the need to merge money and credit
  at all levels so as to obtain liquidity, and we emphasize that this
  structure is made \CJE{clear through the use of} oriented graphs
  which are mainly organized in a tree structure.

%DONE

\subsection{From commodity to fiat money}

%Before we go any further, a comment is in order about the history of
%monetary systems. 
The story we presented for the integration of banks
in a network is only a pedagogical story, and is by no means {\it
  history}. 
%First it cannot be history, because there is
%absolutely no reason for the different countries and epochs to have
%followed the same paths. One story cannot be history.
%Furthermore, questioning the story of monetary arrangements is nearly
%equivalent to questioning the origin of money and currencies. 
The presentation adopted here would imply, if taken at face value, that there was first barter,
then in a \CJE{refinement}, there was exchange of commodities, and only
after came a mild form of fiat money as a claim on commodities (gold)
toward banks. 
%This story implies that there is a positive historic movement toward liquidity,
%but where the value of claims is backed by gold. 
This point of view is essentially the {\it metallist}  point of view or the {\it commodity
  theory of money}. If this might have been true for some societies at some times, there is no proof that
this has been a general feature. And as we have shown in
\S~\ref{SecBankRun}, it is inconsistent when the banking system allows
for the creation of credits.

It is argued by~\citet{Graeber2011} that this is rather the exception in history, and that the most
generic type of money is credit money which finds its origins in the quantification of moral debts in units of
accounts. The gold standard money appears in that perspective as an
exception driven by the industrial revolution, and ended in 1971.
These general ideas are rooted in the {\it chartalist} description of the origin of money, where the UoA for the
debts is actually set by the state. It dates back from Mitchell-Innes~\citep{Innes1913} who emphasized
that money is a standard of deferred payment. As the state spends in a
chosen and arbitrary UoA, it also sets that this unit
should be used for debt repayment, as all individual debts toward society which take the form of taxes have to be redeemed in
that state UoA. In that case, this pure fiat money does
not derive its value from the market as a commodity, but is rather credit money whose value is
initiated by the sovereign states. These ideas have been further developed by~\citet{Lerner} and \citet{Knapp} and are currently
revived under the name {\it neo-chartalism}~\citep{Wray2004}. 

In order to shift from a commodity money to a pure fiat money we just
need the central bank to give up on its commitment to convert its
IOUs into gold. %There is a vault full of gold, but the key has been lost. 
In that case, any customer can ask for its assets to be converted into
central bank's IOUs, usually paper money, but not into gold. The
{\it chartalist} theory of money argues that everything would behave
as usual as long as the state continues to tax in that
UoA. It thus argues that taxation is a {\it sufficient condition} to
impose that the state IOUs are used for debt repayments. The true amount of commodity inside the state-issued coins does not matter, as long as the state has \CJE{a} monopoly on
coinage and the ability to tax in this unit. 
%The ultimate form of coin debasement is thus paper money,  for which
%there is no more hypocrisy and is made out valueless paper.
The power of states lies in their ability to tax and neo-chartalists
thus refer to such a fiat money system as a {\it sovereign currency}. 

\citet{Forstater2003} [see also~\citet{ClassicalEconophysics}] has
recently explored monetization in colonies by colonizing countries,
and showed that it was performed exactly following that logic. In many
cases, the colonizing countries spent their currency in the colonies, and enforced its use by
promising to tax in that same currency. When the US landed in Europe
on 6 June 1944, they wanted to impose such a chartalist monetization,
and it was strongly opposed by the French government in exile,  precisely on the ground that this would amount to colonization~\citep{deGaulle} and not liberation.

%This controversy about the nature of money is extremely \CJE{charged} in
%the macroeconomic debate. But as we have argued already, there is no
%unique theory of money. 

%%There have been societies at some given time
%%that have evolved using a commodity money, and other societies at
%%other times that have developed a chartalist system. In general, one
%%could say that whenever a state collapses, the corresponding society
%%reverts to a commodity money, whereas when a society self-organizes
%%around a strong central power, it shifts toward a chartalist system
%%based on debt money initiated by the state
%%ability to tax~\citep{Graeber2011}.  
Given that the gold convertibility of the Bretton-Woods agreement has been abandoned in 1971, there is no
doubt that the current international system is a chartalist system.
%If this is still debated, it is only because this can be considered
%as being rather recent from a historical point of view. 
%We explore in greater details the debt nature of money in \S~\ref{SecDebt}.
In the next part, we  detail the role of the state in the
monetary arrangements based on such pure fiat money, and we discuss budgetary tools.

%DONE

\section{State representation}\label{SecState}

%Until now we have mainly ignored the presence of a sovereign state, as we have only considered its ability to organize the banks around a central bank and
%to organize the conditions of borrowing (the monetary policy). We have completely ignored its ability to borrow, tax and spend, that is
%we have ignored the chartalist nature of money in our graphs. 

\subsection{Borrowing and spending: Budgetary policy}

\subsubsection{Borrowing methods}\label{SecBoCen}\label{SecBorrowBanks}

The state has a bank account at the central bank, called the {\it
  Treasury}, and it feeds it by \CJE{requiring citizens to pay taxes}. Any taxpayer, creates a new loop when he is asked to pay the Treasury, and
by the usual removal of the loop described in~\S~\ref{SecPay}, the Treasury account at the
central bank increases, whereas the customer's account and the
account of its bank at the central bank decrease \CJE{accordingly}. 
It is formally equivalent to taxpayers converting their assets into
central bank paper money and handing them over to the state. From a
chartalist perspective, this taxation drains central bank IOUs out of
the private sector, enforcing the need to use and hold them.
%Asking to pay tax liabilities in central bank IOUs is certainly a sufficient condition for its
%acceptance as the central means of payment since the customers need to
%ensure that they hold enough of these IOUs, either directly from the central bank in paper
%bank notes, or indirectly by holding IOUs of commercial banks, to be
%able to pay their taxes.

By contrast, public spending goes from the Treasury account to
individuals. For sure, the state can always tax more than it
spends. However, if the state wants to spend more than what it taxes,
it needs to create a public debt which can take various forms. 
\begin{itemize}
\item {\it Money printing}: This is the easiest possibility. 
%At first, one might think that this is also the worst. 
It consists in increasing unilaterally the amount on the Treasury
account. Since there is no more convertibility this is technically possible, because the IOUs of the central bank are not exchangeable
with anything. 
%If there is still convertibility, this might be equivalent to a debasement of all IOUs. But we have already
%seen that as soon as we allow for debt, and this debt is encapsulated
%into the monetary system, convertibility should be abandoned.
 
\item {\it Directly borrowing from the central bank}: Another possibility would be not to print money, that is to put it on
the Treasury account out of thin air, but to lend it to the Treasury. This is
exactly as when money is printed, except that now the state also
issues a Treasury bond (T-bond) to the central bank. Such a situation can only really work
if both the Treasury and the central bank obey the same sovereign
power, so that the state actually controls both sides of the deal, and
it is only a formal arrangement. If this is the case, then it can
borrow as much as it wants. This is fundamentally different from a country
borrowing in a foreign currency, and it is the essence of a sovereign
currency. And when it needs to reimburse, it can just borrow again
what is needed. 
%This would be equivalent to writing in red ink on the
%bond issued ten years earlier that it should be extended another ten
%years. 
Eventually, as the bonds remain and are extended or replaced by similar bonds, this is equivalent to money printing. 
The process of directly borrowing from the central bank is depicted in
the left plot of Fig.~\ref{fig12A}. Initially the net money of
customers and bankers is $M$ and the Treasury borrows $S $ to the central
bank in the left graph. The right graph depicts the result after
spending the amount borrowed, and the total net money of customers and
bankers has increased to $M+S$. This
increase can be traced to the central bank as we see that the
liabilities of the central bank toward the commercial banks now include the spending of the Treasury. Everything happens as if the T-bonds were the new gold
from which the central bank originates its liabilities. If we consider
the consolidated state, which would be the Treasury and the central
bank merged, there is creation of total net money.

\begin{figure}[!htb]
\includegraphics[height=0.35\textwidth]{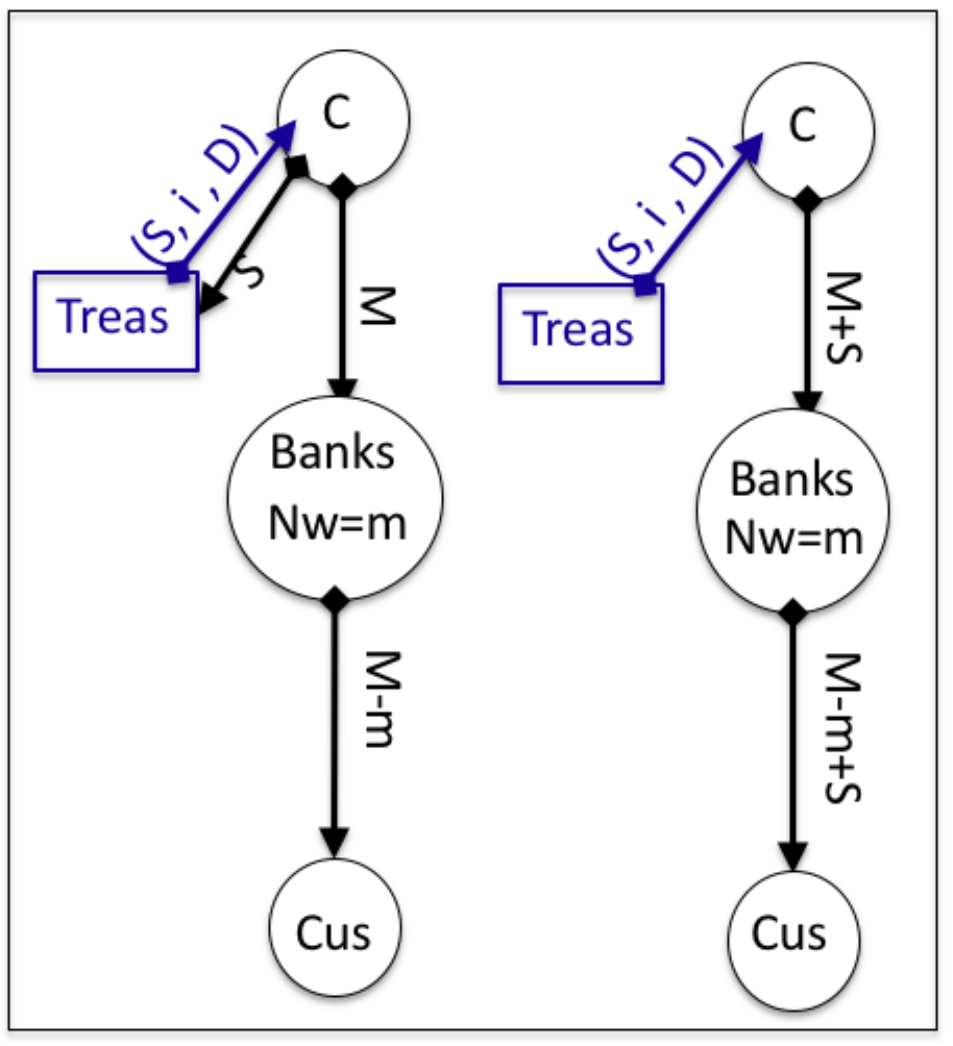}
\includegraphics[height=0.35\textwidth]{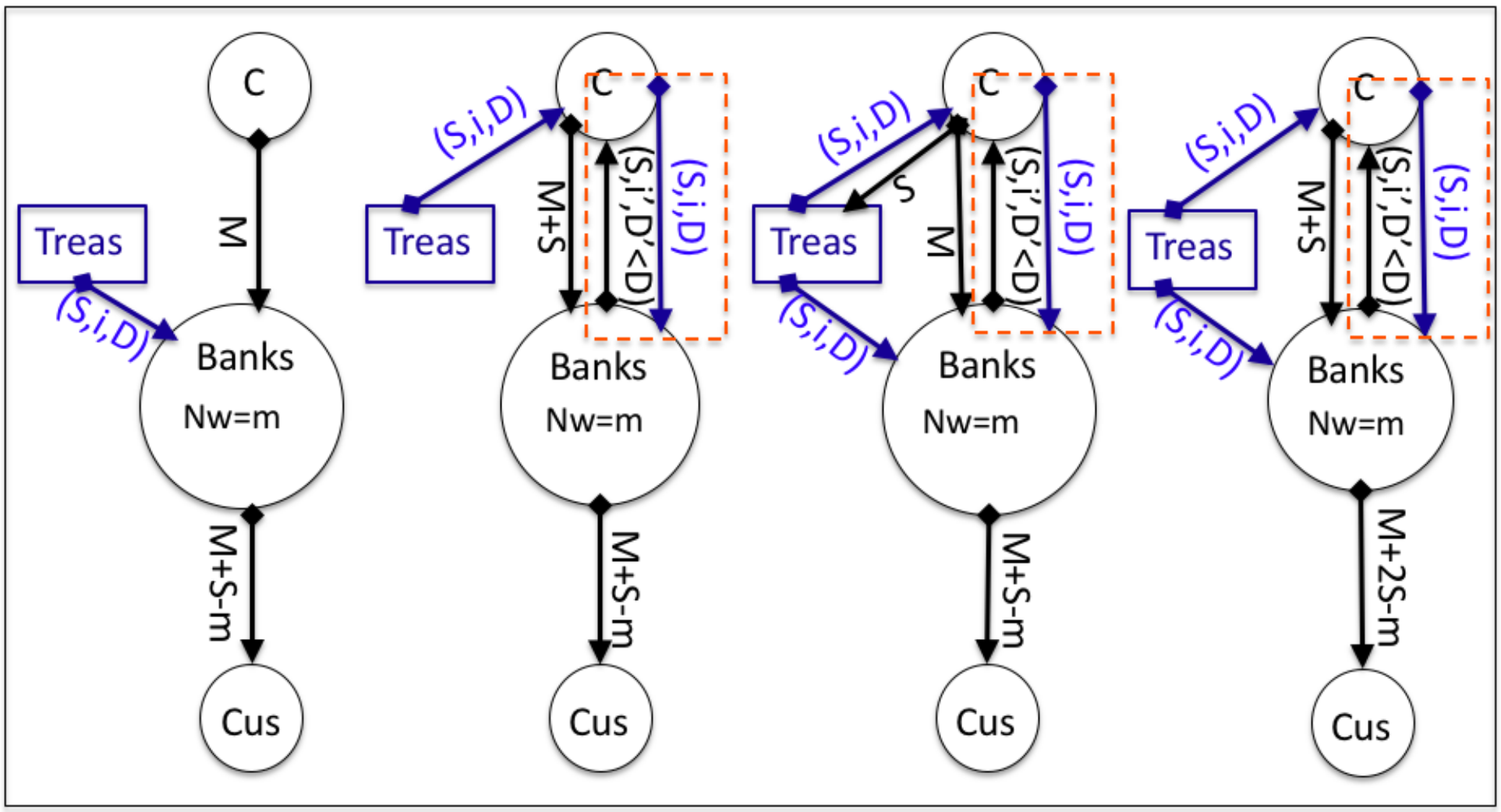}
\caption{{\it Left}: The Treasury directly borrows from the central
  bank and spends. {\it  Right}: The Treasury borrows from the banks,
  after they have themselves borrowed central bank money with a
  repurchase agreement, and then spends.
}
\label{fig12A}
\end{figure}

\item {\it Borrowing from banks}: In this last case, the state is
  forbidden to borrow directly from the central bank and needs to
  borrow from commercial banks. 
%This process goes through essentially  the same steps as when an
%individual borrows from another. 
The result is that in exchange of a payment to the Treasury,
the state will issue a bond, that is an IOU toward those who have paid
the Treasury. It can be individuals, but this is unusual. Commercial
bankers borrow from the central bank what they want to lend to the Treasury if they do not have enough reserves at the central bank. This latter method is slightly more complicated as banks need to
provide a collateral, for instance a previously owned T-bond,
in a repurchase agreement and this is illustrated in the right plot of
Fig.~\ref{fig12A}. In the first graph, we assume that the banks already have a T-bond in their
possession, and then in the second graph they use it to engage in a
repurchase agreement so as to borrow central bank money. At first a commercial bank sells a bond to the
central bank against a smaller amount of central bank money, and then it buys it
back at its value, the difference between the two amounts being
effectively an interest paid. The maturity of the loan is smaller and for this to be interesting, the interest rate needs also to be smaller than the
interest served on T-bonds. The collateral is there to make sure the
bank will repay the loan in one form or another. Either it repays by
buying back the collateral, or it defaults and the central bank keeps
it. The result of the repurchase agreement is that instead of \CJE{directly
holding} the collateral, the commercial bank holds a claim on it, and we highlighted it
by the orange dashed box. In the third graph the commercial bank
transfers the amount borrowed previously toward the Treasury in exchange of
a newly issued T-bond. In the final graph the Treasury has spent the
amount borrowed. Note that the Treasury could have also decided to
open an account at the commercial banks so that the amount borrowed would never have been transferred to Treasury, but after spending the result is the same. 

When commercial banks lend to the Treasury, the amount of IOUs possessed by the
bankers remains constant as can be seen by comparing the first and the
third graph of Fig.~\ref{fig12A}, right plot. \CJE{However,} the nature has shifted from central bank
IOUs which promise nothing, to Treasury IOUs which promise central
bank IOUs in the future with interests. This new T-bond held in the banks net worth
can then in turn be used as a collateral for further borrowing at the
central bank, and further lending to the Treasury, without any
theoretical limit.  Note also that to obtain a clear visualization, we have
consolidated the commercial bank sector and also the customers, but in reality one bank is
involved when it lends to the Treasury, and other banks are
subsequently involved through interbank payments when the Treasury spends in the various parts
of the public sector.
\end{itemize}

%DONE

\subsubsection{Net money and financial wealth} \label{SecNetMoney}

As a result of government borrowing and spending, the total net money of customers is increased by the amount
spent by the Treasury, $S$ in both cases depicted in Fig.~\ref{fig12A}. In the case where the Treasury borrows directly
from the central bank, this increase of financial wealth is reflected
by an increase of liabilities of the central bank toward the commercial
banks. Effectively, the central bank converts the IOUs of the Treasury
into its own IOUs. However, \CJE{if} the government needs to
borrow from commercial banks, the increase of financial wealth has
a double origin. Some comes from the liabilities of the central bank, and some
comes from the liabilities of Treasury. Effectively, the commercial
banks convert the IOUs of the Treasury to their own IOUs which are then
redeemable in central bank money. We conclude that when the
Treasury borrows directly from the central bank, the T-bonds are
converted into net money by the central bank, whereas when the
Treasury borrows from a commercial bank, this conversion is made by the
commercial banks. 

Consolidating the Treasury and the central bank together helps to
interpret this. The total net money of customers has increased by the amount spent by the state. In the first
case the Treasury IOUs are internal to the consolidated state, whereas by
borrowing to banks, the T-bonds now leak out of the
consolidated state sector as they are assets of the banks. In both
cases the net worth of the consolidated sector of banks and customers
has increased, and a natural interpretation is that the net money or
the financial wealth of the public sector has increased.
However, a different consolidation is possible and one might prefer to
consolidate the banks, the customers and the Treasury together in what
we consider as the public sector. In that  case we observe
that the total net worth is unchanged. We realize \CJE{in} this
example that the interpretation of a given monetary system \CJE{crucially depends} on the
consolidations we perform so as to simplify and describe it. We
postpone a detailed discussion about choices of consolidations to \S~\ref{SecStateConsolidation} and \ref{SecRicardo}.

%DONE

\subsection{Open market operations} \label{SecMonetary}

It can be argued that it would be much more democratic if Treasuries were allowed to borrow directly from their central bank.
By electing a government with such a program, we would know what deficit it
intends to run and thus how much it will be willing to
print, which in the long run is a debate about the possible level of
inflation. Instead, it has been argued that decisions made on democratic grounds
might be unstable as they are affected by elections. 
However, the independence of central banks also serves the
interest of commercial bankers as we detail now. 

In practice, the central bank buys and sells bonds in open market
operations\footnote{The case of the European Central
Bank is more complicated since it is not allowed to buy T-bonds in normal
times, but it has engaged recently in such programs to stabilize the eurozone.}. At least it is always doing
so with short term T-bonds as part of the conventional monetary
policy, and it might decide sometimes to do it as well with longer maturity T-bonds as part of the unconventional
monetary policy. This blurs the line between a model where the
central bank directly finances the Treasury, and a model where this is
done by commercial banks since they result in the same final
situation. Indeed, before an open market operation the Treasury owes central bank
money to a commercial bank, and in the final situation it owes it to
the central bank itself, and the central bank money held by the
commercial bank has been accordingly increased.

The commercial bank \CJE{has agreed to dispose of} an IOU which bears
interest, in exchange \CJE{for} a central bank IOU which bears no interest. 
However the Treasury will never default on its debt, because the state
also runs the central bank which can buy an infinite amount of
T-bonds. Said differently, if the interest rates for short term
T-bonds start to increase as the commercial banks become more and more
reluctant to buy these, the central bank needs to buy as many short
term bonds as necessary to ensure the short term interest rates on T-bonds remain at the targeted level. 
By using these open market operations, a sovereign state running a
sovereign currency has the means to ensure that the banks are always
willing to buy T-bonds, whatever the deficit is.
%When Japan entered deflation in 1990, its public debt kept
%increasing, without causing any problem to the monetary
%structure. Indeed it has reached and overtaken $200\%$ of GDP,  but since all of this
%debt is in the national currency, it did not raise any problem. The
%only problem it could have generated at some point was that it could have fostered inflation,
%but that is precisely what Japan needed and wanted to achieve.

However, this system has a drawback. First when the
commercial bank bought the T-bond, it asks for interests rates
which are at least slightly higher than the interest rate at which
they can borrow from the central bank, and make a profit on the
difference. As the interest rates departed from the target chosen by the
central bank, the latter bought short-term bonds to prevent the short-term rate from increasing. In order to convince a
commercial bank to get rid of a financial instrument which is not
risky and which bears interest, the only solution is to pay more than
the current value of the bond, which amounts to a decrease of the
interest rate on those bonds. The bank thus makes an immediate profit
instead of a larger profit later. 
%It has to pretend very hard that it does not want to sell its bond to the central bank, to make sure the
%difference between the face value and the value paid by the central
%bank is interesting enough. 
This difference goes directly into the net worth of the banker and amounts to money creation.

To conclude, we reach the same \CJE{situation where} the Treasury had
\CJE{directly sold} its bond to the central bank, except that now we have increased the net worth of the bankers. By first selling the
bonds to the commercial banks, instead of selling directly to the
central bank, the bankers realize a small profit, that is they benefit
directly from the operational constraint that the Treasury cannot
directly borrow to the central bank.
%But this profit is an immediate and easy one. So they have on one side to
%pretend they do not like when the Treasury goes into debt, so as to be
%able to ask for the highest possible interest rate, and secretly enjoy
%it since either they make a profit when it falls due, or even
%better immediately if the central bank buys the bonds to control
%the interest rates. 

%In sovereign states, where the central bank obeys the logic of the
%government, the deficits can be huge and still the interest rates
%paid on short term bonds are indirectly controlled by the central bank as it buys
%a huge amount of what is issued toward commercial banks. Actually,
%the central bank must buy whatever is necessary to control the interest rates of the short term bonds, as part of the monetary policy.

The commercial banks will always end up with a part of their assets
denominated directly in central bank money, which bears no interest,
and T-bonds, which bear interest. If we adopt a consolidated
state point of view, where we merge the Treasury and the central bank,
then the commercial banks have two types of accounts. Deposits which
bear no interest, and saving accounts which generate interest, just
like \CJE{standard customers at their bank}. In order to control the interest rate, the
consolidated state shifts the amounts from the interest-less to the interest-bearing account and
vice-versa. 

%DONE

\subsection{State consolidation}\label{SecStateConsolidation}

This point of view of consolidating the Treasury and the central bank
is usually criticized because the central bank is supposedly fully independent.
On the contrary, we can argue that it is also a creature of the state and independence is only \CJE{superficial} since it is decided by the state itself through legal
dispositions. Furthermore, it must react automatically with open
market operations whenever a new public debt is issued to control
interest rates, so that it serves the interest of the Treasury
and thus the state. 

We must recall that there are two types of short-term interest rates that the central bank can control:
\begin{enumerate}
\item the interbank rate (called the Fed funds rate in the US), which is the
  rate at which the banks can borrow with collaterals at the central
  bank. This is depicted in the orange dashed box of Fig.~\ref{fig12A} (right plot). It is
  necessarily capped by the discount rate which is the maximum rate at
  which they can borrow, and above the rate paid on
  commercial bank deposits, if it exists;
\item the short term bonds rate which is affected by open market operations.
\end{enumerate}
If a bank decides to control the short-term bonds rate, then when
conducting outright purchases of short term T-bonds, it increases the central
bank liabilities (central bank money), which decreases the
interbank rate. So controlling the short-term bonds rate implies to control also
the interbank rate.

However, a central bank can decide to control only the interbank
rate and not the rate on short-term bonds. For this, it will only conduct repurchase operations, or
collateralized loans to commercial banks. This would affect the rate
at which commercial banks can borrow, but in general it will not affect directly the
rate of the short term T-bonds. Indeed, the Treasury would then
borrow like any customer, and banks are free to set the conditions. Nearly all developed countries, except the Eurozone, control both
rates and run fully {\it sovereign currencies}. However, the
European Central Bank (ECB) is in general unable to control interest rates on T-bonds. This is why the ECB performs only repurchase
agreements to control the interbank rate, but does not control the
short-term rate on bonds via outright purchases of T-bonds,
contrary to the Fed monetary policy. As there are several
types of bonds, each one of them issued by a different government, the
ECB cannot decide which one to buy as it would amount to a form of
financial solidarity between the various European states, and this is
intentionally avoided in the European Union construction. As the various Treasuries are not helped by the ECB to issue low
interest bonds, everything \CJE{occurs} as if they were borrowing in a foreign
currency, where the interest rates are set by the bankers, just like
for any standard customer. The bankers lending to the Eurozone
Treasuries decide what should be the markup rate, that is the difference between the rate at which they
borrow, which is effectively the interbank rate set by the central bank, and the rate at
which the lend to the various Treasuries. On the contrary, the United
Kingdom (UK) or the USA are \CJE{configured so that their} central bank buys whatever is necessary to
control the interest rate on bonds, and things are as if the Treasury was
borrowing from the central bank, except for the small profit made by
bankers due to their intermediation. 

To conclude, if the Treasury borrows directly from the central bank, it
makes sense to consolidate the Treasury and the central bank. If it
borrows from commercial banks, but the central bank controls the short
term interest rates on the bonds, the effective theory is nearly the
same, and it also makes sense to consolidate the Treasury and the
central bank.
%We must also mention that so far we have considered only
%short term bonds. First, the Treasury could perfectly issue short term bonds
%only. That would certainly work fine if the Treasury was borrowing
%directly from the central bank. And if the Treasury needs to borrow to
%commercial banks, it would be sure to pay the standard rate as the
%central bank needs to control the interest rate on short term bonds
%and must be prepared to buy whatever is needed to achieve this goal.
%But it is a reality that Treasuries also issue longer term
%bonds. Since the commercial bankers need to protect themselves from
%the risk of higher short term interest rates, which is the rate at
%which they borrow, there is necessarily a markup rate and long term
%bonds always have a higher interest rate. But this might precisely be what the
%government could try to achieve. Indeed the long term bonds are not
%repaid at once at the end of their maturity, but instead the interests
%are paid in coupons gradually, with the principal repaid only at the
%end. So by buying these, the banks have a way to predict and control a substantial
%part of their future cash flows. Issuing bonds of various maturities is thus a way
%to stabilize the banking sector and to smooth the issuing of bonds.
However, if like in the Eurozone the central bank stops
controlling the interest rate on bonds but controls only the interbank
rate, the Treasury is treated like a standard customer, as it is
treated by the central bank as if it was a foreign Treasury. 
%We say that this is not a sovereign currency. 
%Note that in that case, the
%governments might find it more convenient to issue only long
%term T-bonds, so as to postpone the repayment after the next
%general election. 
The consolidation might still be possible formally, since any
consolidation can be made as it is just an accounting simplification, but it hides some
salient features. Sometimes commercial banks would add only a
small markup rate to the T-bonds and one would not see the
difference, but in case of crisis the markup rates can start to be
huge, on all maturities, even reaching the point where commercial
banks stop buying T-bonds as in the recent Eurozone debt crisis. 

%DONE

\subsection{Monetary policy}

\CJE{Monetary policy tools} are very complex and \CJE{can vary greatly} from one country to another. It is by no means the goal
of this paper to present them all. The main conventional tools are the control of
short term interest rates, and the reserves requirements.

\begin{itemize}
\item Conventional monetary policy is about controlling the short term
  interest rates. However, controlling the interbank rate is not sufficient to control the rate at which customers borrow. Indeed, when commercial
banks lend, they need to apply at least the interbank rate as it is the rate at which they need to borrow when the loan is used to
pay outside of the commercial bank, or when a part of the amount
loaned is transformed into cash, that is into central bank money.
But they also need to apply a markup rate to this basic interbank rate at
which they borrow for several reasons. First, the borrowers might default and
the bank needs to make sure that it generates enough profits from
non-defaulting loans to compensate for the defaulting ones
%\footnote{When
%  the loan is an investment loan, such as when customers buy houses,
%  the investment in real assets is also used as a guarantee. If the borrower fails
%  to repay the loan, the property of the real asset, e.g. the house,
 % is transferred to the bank.}. 
Second commercial banks need to make a profit to cover their running costs as they need
to pay at least the salaries of their employees. Finally, since they
have borrowed on a short term basis, but they lend on a longer term
basis, they need to have a security margin in case the central bank
increases the interbank rate. For all these reasons, the effective
rate at which the economy is functioning is different from the basic
interbank rate chosen by the central bank. If the central bank wants to foster
credit with low interest rates, it is as important to set a low interbank rate as
to communicate the fact that this interest rate shall remain low, so
as to decrease as much as possible the markup rate. Finally, we must
stress that the interest rate is not the only criteria to ask for a credit,
as decisions are made on much more fundamental economic grounds. Even
if the markup rate remains constant, the interest rate set by the central bank is only an indirect tool to control credit and thus the total money.

\item \CJE{Reserve requirements for commercial banks is the second most common monetary tool.} In the theory of the money multiplier, the
  central money held by banks should be a fraction of its liabilities
  toward its customers, and this fraction should be set by the
  monetary authorities. It is then assumed that by controlling the
  amount of central bank money held by the banks, and by fixing the
  ratio, the central bank could control the total credit, and thus
  the total money. But this tool cannot be efficient, because the
  amount of central bank money held by banks is not exogenously set.
Indeed what is counted as a reserve for a commercial  bank is not its net central money, but its central money. So if the bank does not have enough central money reserves, for instance
  because it has granted too many loans, it can borrow the reserves needed at most at
  the discount rate, and more probably at the interbank rate. 
When the bank does so, at the same time it receives central bank money
and increases its reserves to comply with its legal obligation, but it
owes it as well at a later date, and this does not count negatively in
the reserves. We see that the difference between net central bank money and central bank money is very important. The
central bank can control its net liabilities toward the banks, e.g. by
performing outright purchases of T-bonds, but it cannot control
its liabilities, as these are endogenously determined by the needs of
the commercial banks. In practice, banks lend whenever they think it
is profitable for them, and if they fail to meet their reserve
requirements at the end of the day, they just borrow (directly to the
central bank at most at the discount rate or to other banks) what is
needed.  In a few developed countries (Canada, Australia, UK, Sweden, Norway) there are no more fractional
reserves and nothing special happens. As long as the required
reserves are not net reserves, they are entirely useless. 
%Furthermore,
%even if the fractional reserve system was applied to net central
%money held by banks, this would only set an upper limit to credit, but
%it could not increase automatically the amount of credit if it is below the cap.

%In fact the net assets of the banks, that is their net worth which
%includes the capital which has been given in by the stockholder, is
%instead extremely important. These are true reserves which are going to be used whenever a bank suffers losses. It is thus no wonder that after the 2008 financial crisis, the rules for the fractional reserves
%have not been modified, whereas the capital requirements have been radically increased in the third Basel %Accord.
\end{itemize}

\subsection{Should it stay or Schuld it go: the clash}

\begin{figure}[!htb]
\includegraphics[height=0.37\textwidth]{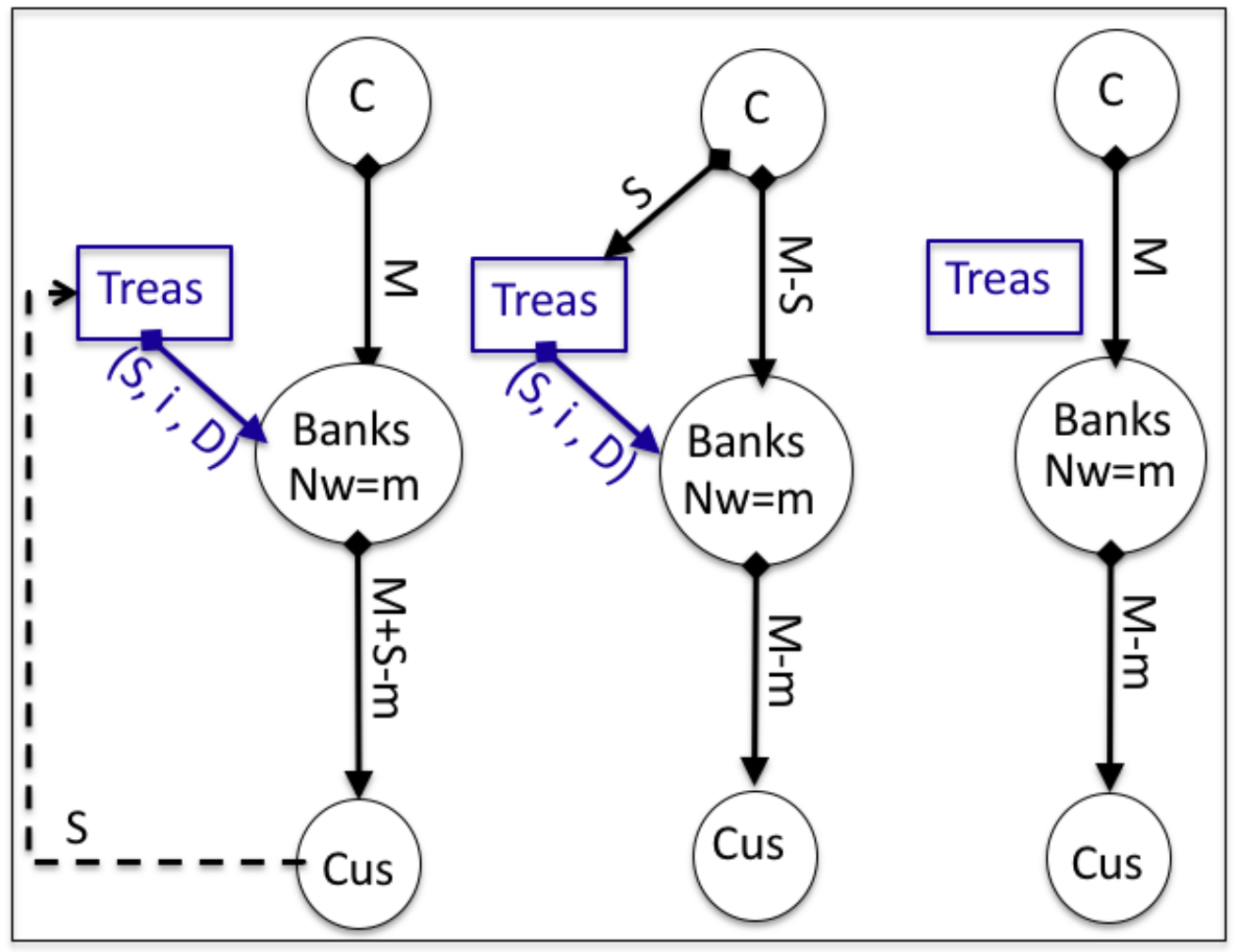}
\includegraphics[height=0.37\textwidth]{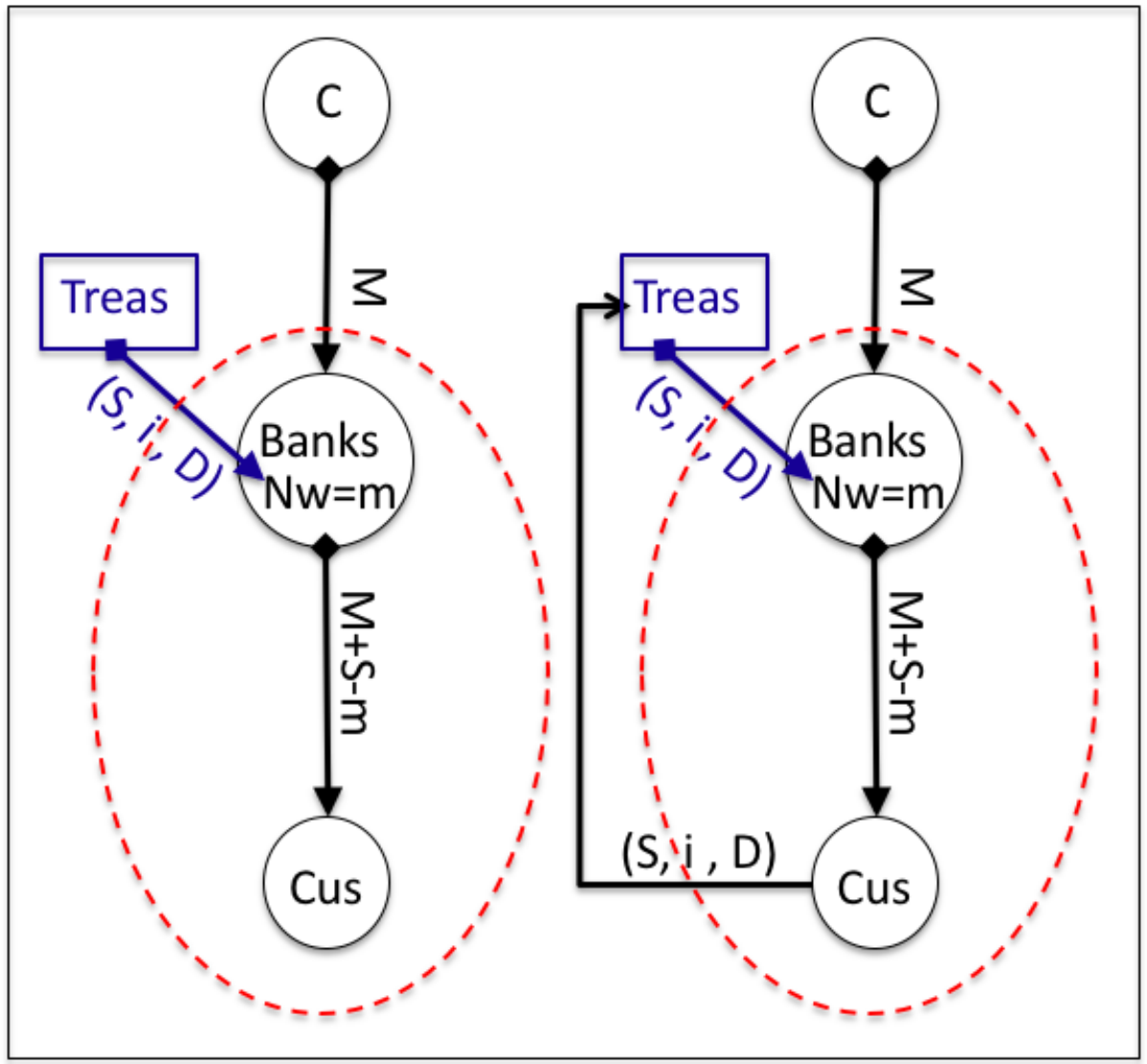}
\caption{{\it Left}: Reduction of the public debt from taxation. {\it
    Right}: The {\it should} and {\it Schuld} points of view on
  the consolidated public sector (shown in the dashed red circle).}
  \label{fig15}
\end{figure}

%We have seen that when the Treasury borrows and spends, it increases the
%financial wealth of customers and bankers. But it is subject to
%interpretation in order to decide if it increases the net money or
%not. It depends how we consider the net worth of bankers.
There is a huge debate on the nature of money and the nature of
public debts. Apart from the fact that T-bonds bear interest, they have an intrinsic difference with central bank
IOUs, which is that they have a maturity, and are thus bound to
disappear through the reimbursement of the Treasury's debt. 
%But we have seen that for the customers, the monetary system has
%resulted in an apparent conversion of the Treasury's IOUs into normal
%bank deposits for customers. 
Let us take a simple case where $20\%$ of
the net deposits is a reflection of the central bank IOUs, and the
rest comes from government debt. 
%The good news is that since this mixing between the different origins has been
%made possible thanks to the banking system, it means that 
By construction the customers \CJE{always have} more assets than the debt of
the state. So the state could always, at least in theory, run a huge
temporary tax on capitals which would reduce the assets of customers
and increase the Treasury account so as to allow debt repayment. \CJE{This
would amount to} one year of GDP since nowadays public debts are of that
order. The payment of the tax is illustrated in the left plot of
Fig.~\ref{fig15}. In the first graph we identify the pending tax
payment, which is then used to credit the Treasury account in the
second graph. In the third graph this is used to reimburse the public
debt. As a result the net money would be reduced by that
amount, and within our example the financial wealth on
deposit and saving accounts would be divided by five. Since 2011, this is
\CJE{essentially} what has been \CJE{implemented} in European countries where it has been decided to
reduce the public debt at any cost. 
%It is essentially lead by the ignorance of the fact that public spending,
%if made from borrowing in the national currency, leads to an increase of the net money of
%customers, and can anyway always be reimbursed. 
%And that is where macroeconomics starts to be a social
%science. 
People \CJE{might} be fine with the reduction of their financial
wealth \CJE{because they believe} that in the long run prices should
decrease, or even changing their saving habits so that it does not
happen. 
%There is in principle nothing wrong and one could imagine that some
%countries might cope with it in happiness. 
But we think that it is more likely that the private sector \CJE{will
adhere to its preexisting savings habits}, meaning that everybody tries to spend less, causing
deflation and possibly after recession. 
%Enlever ca.
So by ignoring this, the goal of debt reduction can be extremely
harmful to the economy. In fact, whenever a state runs a surplus, this never
lasts more than a few years~\citep{Wray1990,Wray2004}, and then recession enters the game to
generate new deficits. If public debts have been reduced when
compared to the GDP during some period of history, they are nearly always constantly
increasing in nominal value, because the total net money of customers
needs to go more or less at the same pace as GDP growth plus inflation
to satisfy the habits. 
%TODO remove uncertainty.
%So to summarize, states run deficits which
%correspond to net money creation for customers, which fosters
%inflation in the long run if it goes much faster than economic (material) growth.

However, all our preconceived ideas run counter proper thinking. For
instance, in German %, which is the country with the worst personal relationship with
%respect to public debt, 
money is \CJE{called} {\it Geld}, a word which is a derivative of {\it  gold}. It
thus carries a meaning \CJE{that} goes beyond the real nature of money where
convertibility has been abandoned. On the contrary, debt is called {\it Schuld}, the same word that
is used for guilt or fault\footnote{As noticed and analyzed in~\citet{Graeber2011}, this is also the case in many ancient
  languages (Sanskrit, Hebrew, Aramaic).}, depending on how we
translate. \CJE{So the naming convention of these basic financial terms may
influence German speakers' economic decision-making}, since as the Germans pronounce the
word {\it debt}, they immediately mean what they should do about it, that is
getting rid of it, as it is morally bad. 
%There is nearly a religious conception of bad
%behind the concept of debt in German, and since they rule Europe, they
%impose their view that, no matter if it triggers deflation and then
%recession, it is fundamental to get rid of this initial sin. 
By drawing the consolidated graph structure, \CJE{we are able to replace
these linguistic biases with a neutral graph}, as only asset/liabilities relations of different types appear.

%DONE

\subsection{Ricardo equivalence against an asymptotic definition of money}\label{SecRicardo}

In the end, the debate is about what we think the debt will become in
the very long run. Indeed, in a system \CJE{that} has abandoned
convertibility, central bank money is an eternal debt. \CJE{In some
  sense,} the central bank owes gold, \CJE{but it owes it without any deadline, given that} it has
abandoned convertibility. It would thus be tempting to define money as a debt \CJE{that} is never
reimbursed. Interestless money in that definition would thus be just
one type of money. \CJE{As an example, let us consider} the paper
debts (Demand Notes) \CJE{that} were issued by the US federal
government during the Civil War. At that time they were issued \CJE{as} debts, in order to finance the war, but they kept being
exchanged and were subsequently considered as currency, without being redeemed into gold, but being used to pay taxes.

The problem with such definition of money is that it is not local in
time. Indeed, we need to know the full future of the debt to deduce if
it was actually money or if it really was a debt which has been
reimbursed.
%\footnote{This construction is thus similar to the definition of horizons in
%General Relativity, where the definition relies on the full future of
%the spacetime solution.}
%The drawback of the definition is thus that we
%do not know what will be the infinite future, since this is
%fundamentally impossible. 
So we have to rely on an imperfect definition,
where we would consider that a debt is money if there is a consensus
among customers that it will never be reimbursed (or reimbursed by
issuing an equivalent debt). This deeply depends on the point of view. For any debt issued by the Treasury, there are thus always
two extreme point of views. \CJE{The first one is held by those} who believe that since states
always run deficits, it is for sure a type of money. We call this
point of view the {\it should} point of view since in this point of
view we should run deficits, at least in pace with growth and expected
inflation. Conversely, for those who think that debts \CJE{should be repaid}, no matter the state
of the future economy, the public debt cannot be considered as money. We can call this point of view the {\it Schuld}
point of view. The debate between {\it should} and {\it Schuld} is
reminiscent of the debate about rational expectations. If the state
issues a new debt, does it mean that there is an actual debt of the
customers toward the state? Those who think that this debt exists no
matter what are in the {\it Schuld} \CJE{camp}, whereas those who ignore such possibility, by arguing that people do not look at
national accounting for the personal wealth, are in the {\it
  should} camp. For the {\it should} point of view, since the state is everybody
it is nobody and lies outside. For the {\it Schuld} point of view, since the state is nobody,
then it means that it is everybody and it lies inside. The {\it Schuld} point of view was first
formulated formally as the  Ricardo-Barro equivalence, according to which
taxpayers exactly anticipate future taxes from current deficits. The
two points of view are summarized in the right plot of
Fig.~\ref{fig15}, the left graph and right graph being respectively
the {\it should} and {\it Schuld} points of view. As
already mentioned in \S~\ref{SecNetMoney}, if the
Ricardo-Barro equivalence is invoked, the financial wealth of
customers and bankers always remains unchanged, even when the Treasury
borrows, and remains $M$, that is the total liabilities of the central
bank, whereas if the equivalence is ignored (as we think it should
be), the financial wealth is increased by public deficits $S$. 
%and is thus $M+S$, where $S$ is the total debt.

It is hard to \CJE{believe that people's decision-making will be influenced
by a} possible debt they owe to the Treasury. We think that in order to find a convincing answer in this
debate we must look at the past behaviour of major western states. And
what we find is that, apart for a few years, they run constant deficits and the
sovereign debts keep growing in nominal values. Only on some
occasions, some governments manage to run a surplus but this never
lasts very long. For the US history, this is certainly the case, where
there was just occasional years of surplus in an ocean of
deficits~\citep{Wray2004}. The need of running deficits to ensure
growth has been advocated by \citet{Minsky1986}, and it is best understood from the
sectoral financial balances perspective which was popularized as a tool of macroeconomic analysis by Wynne Godley~\citep{Godley1999,Godley2000,Godley2007}.
\ifsfb
See \S~\ref{SecSFB} for more details.
\else
\fi

%DONE

\ifmoneyagg
\section{Monetary aggregates}\label{SecMi}

Having qualitatively described the monetary systems and understood their
various mechanisms, it becomes necessary to define monetary aggregates
that allow for more quantitative statements on actual monetary systems.
All monetary aggregates correspond either to the net money of a
consolidated subregion, or to the net financial position between two
subregions, that is the net debt from one region to another one.
Again, since the subregions are numerous, so are the various monetary aggregates.
\CJE{Using just a} few ones to describe the dynamics of the
monetary system reflects our interpretation of these systems.
All aggregates that we now present are illustrated in Fig.~\ref{fig22}.

\begin{itemize}
\item $M_0$. The simplest monetary aggregate is $M_0$ and is composed of the assets
directly held by the customers at the central bank, in the form of
coins and paper money. Whenever a customer deposits money at its
commercial bank, this amount is reduced, and conversely it is
increased when deposits at commercial banks are withdrawn in cash. 

\item $M_B$. If we then include all net deposits held by commercial banks at the
central banks, this defines the monetary basis $M_B$. Its definition
corresponds to the financial position between the central bank and the
public sector.  By construction, it excludes everything  which
is related to the public debt issuing in its definition. Hence, the monetary basis will always vary whenever the central
bank performs open market operations. Finally note that our definition for $M_B$ takes
into account only the net deposits of the commercial banks at the
central bank, and not just the deposits. This is different from the
usual definitions of the monetary basis. With our definition, any central money borrowed at the
discount window does not affect the monetary basis $M_B$, whereas it would
affect the usual definitions of the monetary basis.

\item $M_{\rm net}$. This arbitrariness leads to consolidation of the full aggregate of
customers and bankers $M_{\rm net}$. It counts the
Treasury debts as well, as it includes all types of assets entering
the bankers net worth. Whenever the central bank engages in open
market operations, buying or selling bonds, this monetary aggregate
does not change. It corresponds to the net financial assets held by the
whole economic system and this is what we have called the {\it net money}.
An operation of quantitative easing, which is just a massive outright
purchase, changes $M_B$ but not $M_{\rm net}$. 
%Since the whole system perceives only its net position, a quantitative easing has no
%major effect except a psychological one if widely advertised, and a small effect on
%the profits made by bankers when selling their T-bonds above market
%prices to the central bank. This is where a clear understanding of the
%aggregate under scrutiny, that is of the consolidation which is
%considered, is crucial for the debate. 
If we use the ambiguous word {\it money} for different aggregates, we would unavoidably disagree
on the effect of open market operations. The use of graphs in
representing \CJE{this scenario} clarifies the debate.

\item $M_{\rm CB}$. Whenever the Treasury spends by increasing its
  debt,  $M_{\rm net}$ is increased. This leads to the definition of
  the central bank aggregate $M_{\rm CB}$ which encompasses
everything but the central bank, and this one will remain constant,
whatever the public debt. This aggregate is the reflection of all gold
and foreign reserves of the central bank.

\item $M_i$. \CJE{We can introduce a further complication} if we now decide to exclude
the liabilities of customers, but to include only some assets, in an
attempt to extend $M_B$. This leads to the various definitions
$M_1, M_2,\dots$, that we gather collectively as $M_i$. The more
assets of customers are included (with all liabilities excluded in all
cases), the larger the aggregate $M_i$ is. We thus have a hierarchy $M_1
< M_2<\dots$. Any loan issued by a bank would surely affect some of the $M_i$
as it would increase the assets held by customers. Depending on the
nature of the asset, it would count in some $M_i$ and not in
others. These definitions of the aggregates are a way to estimate the amount of credit made by
commercial banks. More precisely the ratio between the $M_i$ and $M_{\rm
  net}$ can be used the estimate the amount of credit in the origin of
money.
\end{itemize}

\begin{figure}[!htb]
\includegraphics[height=0.40\textwidth]{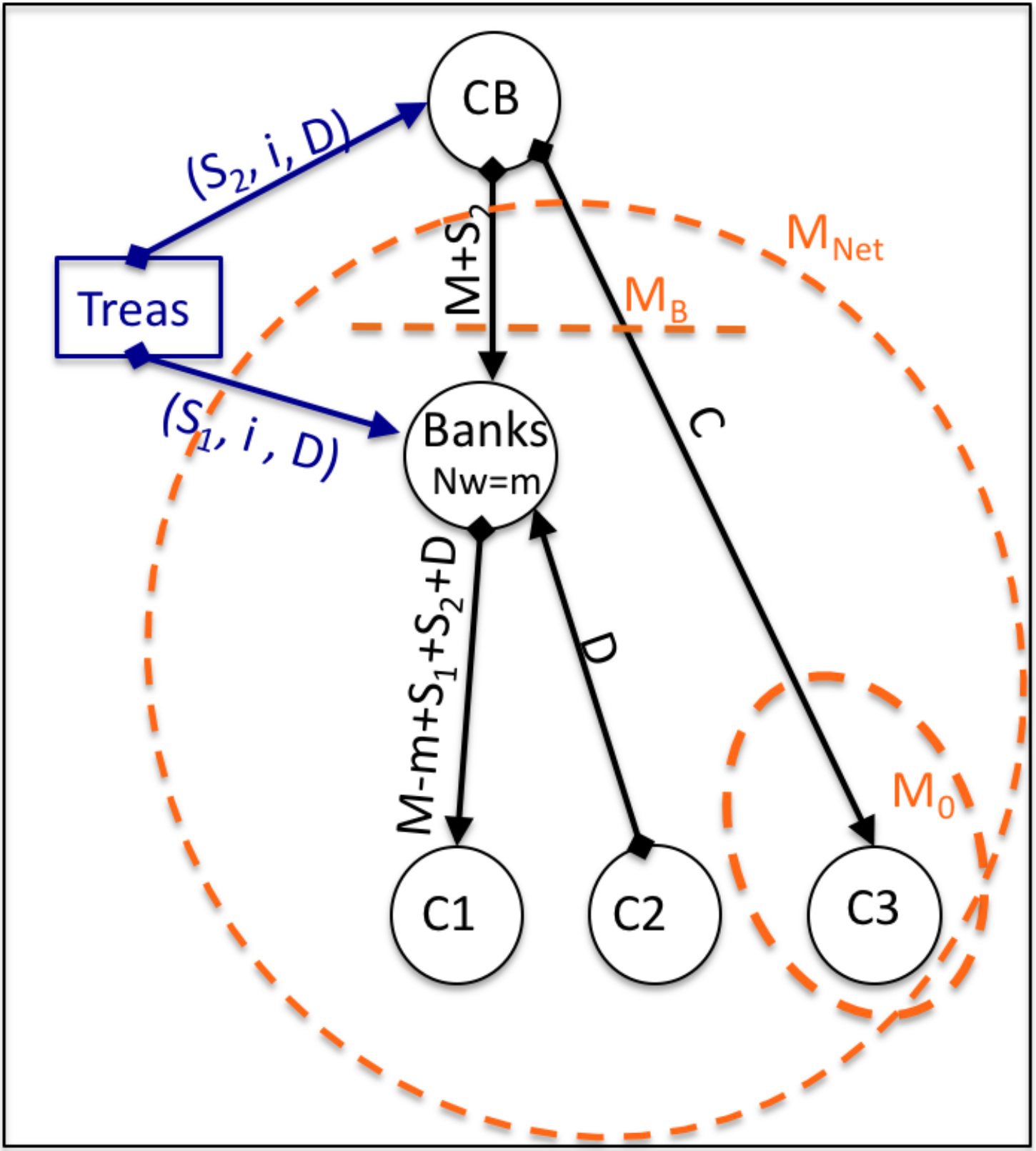}
\includegraphics[height=0.40\textwidth]{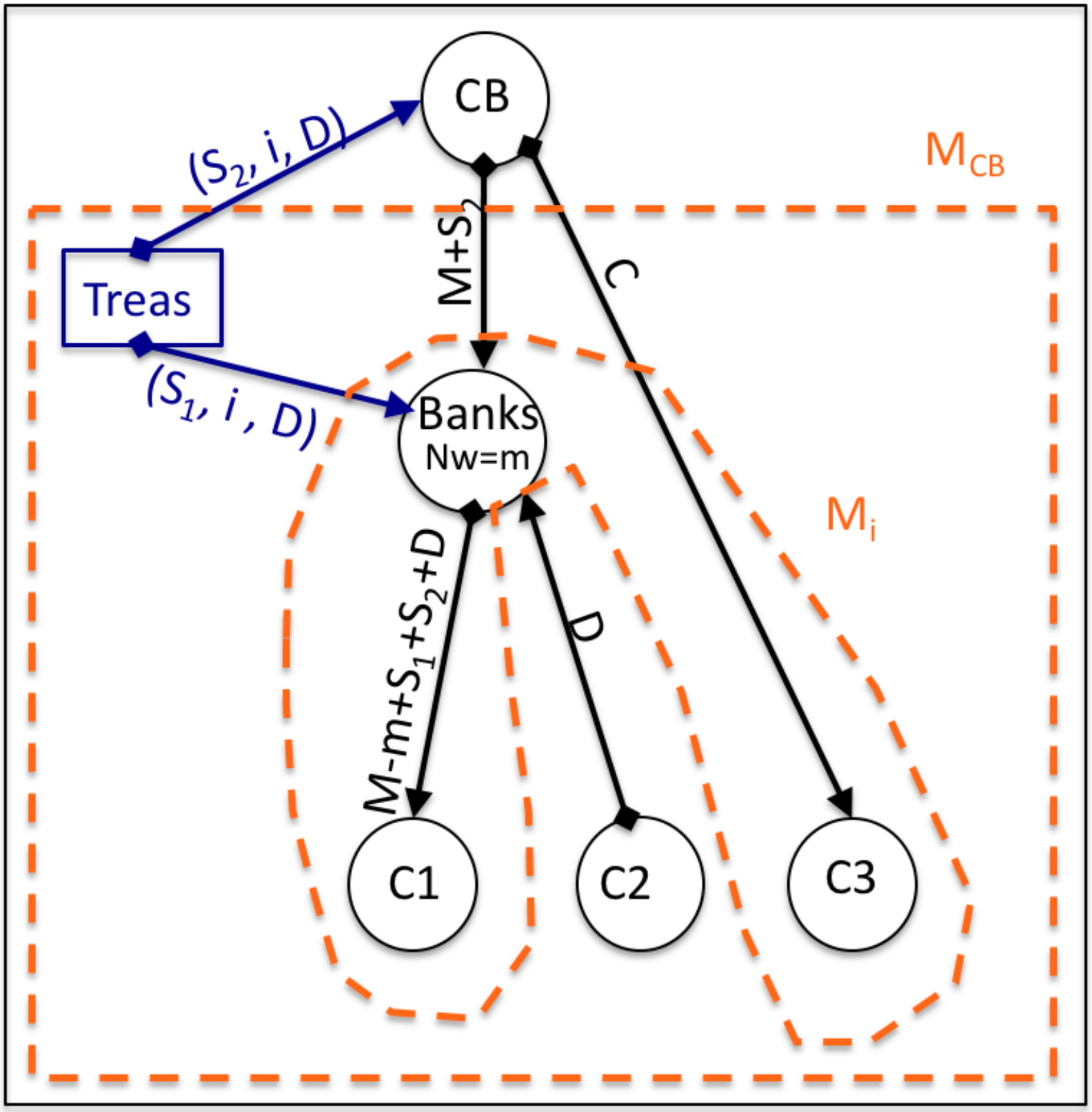}
\caption{Schematic representation of $M_0$, $M_B$, $M_{\rm net}$, $M_i$ and $M_{\rm CB}$.}
  \label{fig22}
\end{figure}
\else
\fi

\section{Foreign currencies and exchange rates}\label{SecFor}

There are inevitably several sovereign states, each with its own currency, that is its own central bank and its
own Treasury, using its own UoA. A complete description of
a general monetary system must thus allow for several currencies and their interactions.

\subsection{Fixed exchange rate system}\label{SecExRate}

We first focus on fixed exchange rates systems \CJE{that}
were more common in the past. In this section we describe the
interactions of various monetary systems, and in the next section we
argue clearly why this can never be maintained.

\begin{figure}[!htb]
\includegraphics[height=0.30\textwidth]{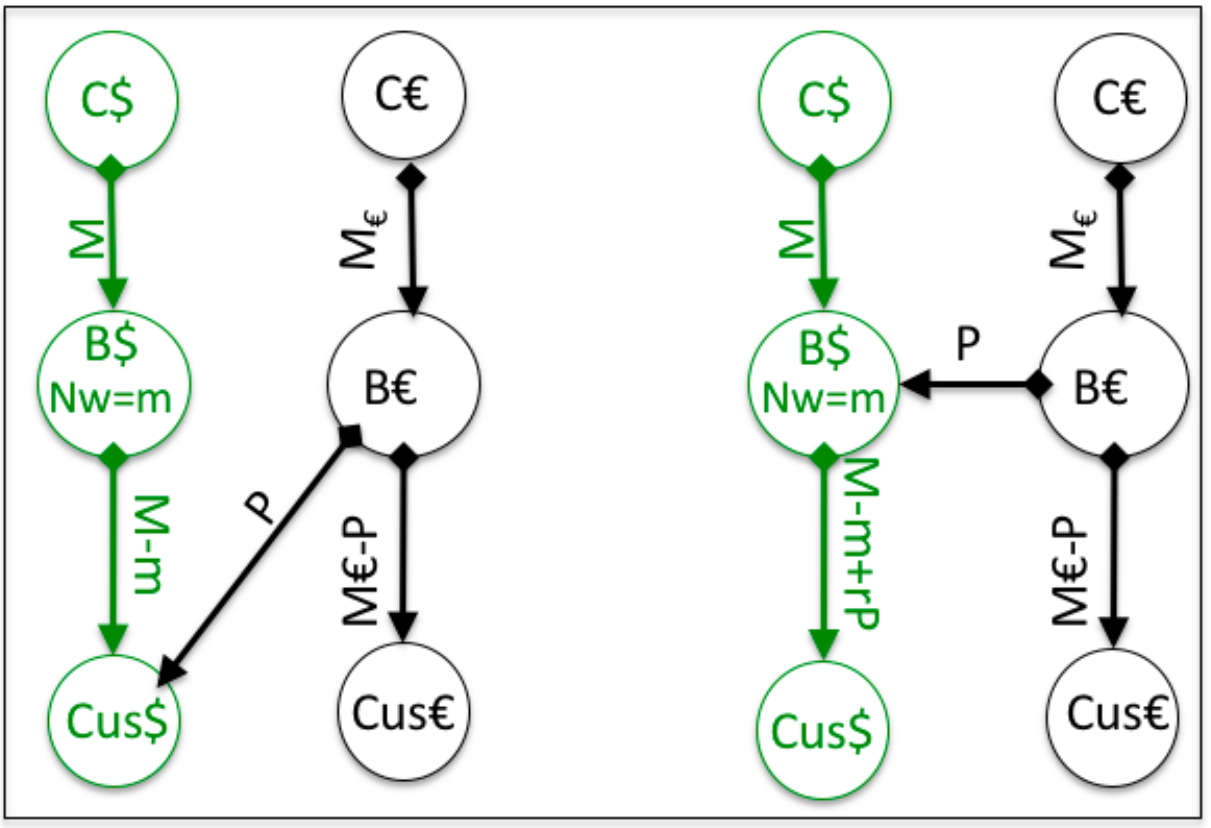}
\includegraphics[height=0.30\textwidth]{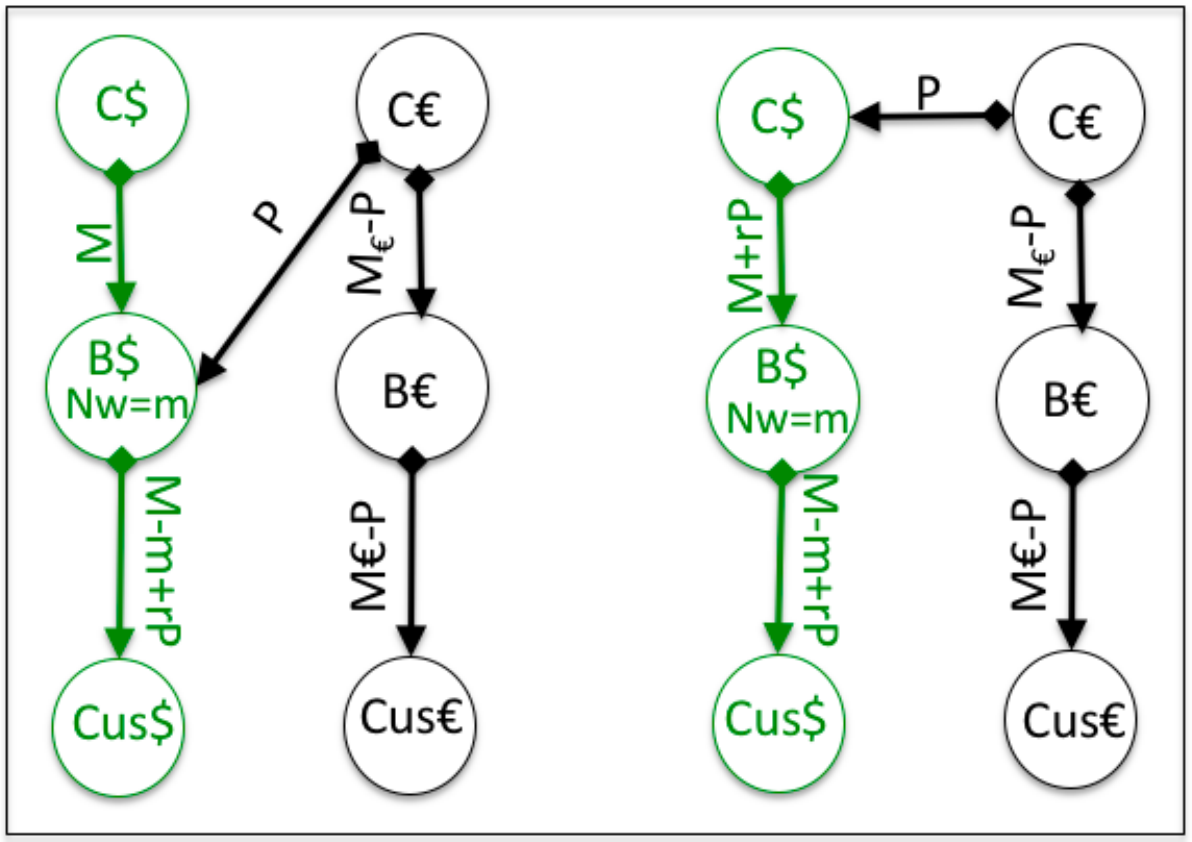}
\caption{The foreign currency held be a customer is
  transferred toward its bank in the left plot and then to the central
  bank in the right plot.}
  \label{fig18}
\end{figure}

Whenever a national citizen, is paid by a foreign citizen in a
foreign currency, he might prefer an asset denominated in his national
currency rather than an asset in a foreign \CJE{currency}.
%There is no universal law behind this, and all possible
%choices are in principle possible. He might find it perfectly fine
%and preferable to hold foreign currencies. 
In that case, he will ask his bank to exchange it for a national
asset as illustrated in the left plot of Fig.~\ref{fig18}.
%In a fixed rate exchange system, the amount which should be traded is
%obvious. 
The commercial bank will take the foreign currency for itself, and will create a deposit denominated in the national
UoA according to the fixed exchange rate. The process might then be repeated
between the commercial bank and the central bank as illustrated in the right plot of Fig.~\ref{fig18}.
%The commercial bank might prefer to hold a liability of its national
%central bank than a liability of a foreign central bank. 
The central bank will take the foreign central bank money as an asset and will increase
the  national central bank money deposit of the commercial
bank. As a final result, it is the national central bank which directly possesses the liability of the foreign central bank, that is which has
a deposit at the foreign central bank. If the whole procedure is reversed and the customers
prefer to hold assets denominated in a foreign currency, this is
usually called {\it capital flight}.

\subsection{Fixed exchange rate limitations}\label{SecPeg}

The resulting situation is extremely similar to what would have happened if the
national citizen had been paid in gold under a regime of convertibility,
and if at every stage the gold had been passed to the higher
level (from the citizen to its bank and then to the central
bank). 
%The resulting situation is indeed that the total net money of the national sector has been increased by $rP$, where $P$ is the
%initial foreign payment, and $r$ is the rate of conversion to the
%national UoA. 
%It is because of this analogy that commodities and foreign currencies held by central bank are treated
%essentially on the same footing. 
In a fixed exchange rate regime, foreign currencies are treated like gold in a regime of
convertibility. If payments are made in gold or in  foreign
currencies, they can be passed over to the central bank \CJE{that} creates
the corresponding net money. Conversely when gold or foreign
currencies are \CJE{sought}, the central bank reverts the situation and
detaches the asset from its balance sheet, but in the process it also
reduces its liabilities in central bank money, effectively reducing
the net national money.

\CJE{As demonstrated in \S~\ref{SecBankRun}, in a regime of convertibility
between money and gold,} the system cannot hold because of
credit creation. The problem is exactly the same in a regime of fixed
exchange rate. It is always possible to break down the fixed
exchange rate by asking the conversion of national currency into
foreign currency, and given that there is more money than net money,
there is always a way to reach the point at which the central bank
reserves are depleted.

The only possibility for the system to continue would be if the national
central bank had the right to borrow foreign central money to the
foreign central bank. But then the situation
would be rather similar to the case of commercial banks having the
right to borrow from a central bank, and this would ensure that all
liabilities can be cleared at par, that is with a fixed exchange
rate. This was exactly the European situation between 1999 and 2002,
when all eurozone currencies had a fixed exchanged rate with the euro
foreign currency, and all national central banks were granted the right to borrow from the ECB as much as they
needed. Eventually, all national currencies were abandoned and the
system was further integrated into a single currency, as this was just
a technical intermediary situation. But in general a foreign central
bank has no particular interest in granting the right to borrow to the
national central bank.

We realize again that granting the right to go in debt is the key \CJE{to} financial power. At every vertex of the tree structure, granting the
right to go in debt to vertices which are lower is a form of power on
them. The household is afraid that its commercial banker might not grant a
loan, and needs to accept the conditions set by the banker, the
commercial bank is told at the higher level at which rate, under which
conditions, for which maturity, it can borrow central bank money. And
if the national central bank now wants to borrow a foreign currency, it
will have to accept the conditions of the foreign central bank.

If an economy \CJE{fixes} its currency at parity with a foreign
currency, its central bank somehow inserts itself in the tree
structure of the foreign country, and instead of being at the apex of
the system, it is now situated below the foreign central bank and
needs to ask for the permission to go into debt. Even though this type of
power of the foreign central bank onto the national central bank is
formally similar to the power a commercial bank has on its customers, it is
also extremely different.
A central bank possesses a form of power on the national financial
system as it sets the condition of debts in commercial banks and thus
below for national customers/citizens. At least in principle, it is acting for the interest
of the national economy as it is a creature of the state. Instead,
when a foreign central bank grants the right to a national bank to
borrow foreign central bank money, it has absolutely no reason to act
in the interest of the national citizens. 

If a weak economy wants to fix its currency to a strong economy, and does not have a full constitutional right to borrow in this
external currency, as is always the case, there are essentially two possible situations. Either the country ensures that it will never need foreign
currencies to defend the exchange rate. In order to achieve this it
would need  a strong commercial surplus. Or it needs to borrow the
foreign currency, e.g. because of a current account deficit, and needs to
comply with the politics imposed by the foreign country. The alternative
is thus either to work for free (this is the essence of a permanent
commercial surplus for which the dominant foreign countries pays simply by
public deficits) or to be told how to work and what to sell (structural reforms
imposed by creditors). 

\section*{Conclusion}

We have argued that graphs for financial stocks are natural tools to describe and discuss the
theories of money, because \CJE{every} monetary arrangement is inherently a
graph. By connecting together the balance sheets in a graph rather than explicitly writing one after
another the balance sheets of each institution [e.g. in the figures
of~\citet{BoEMoney} or in the balance sheets explanations
of~\citet{Keen2014}], we found that \CJE{a} hierarchical nature
\CJE{expressly} appears. We advocate \CJE{using} graphs
whenever monetary systems are discussed, as they allow for a more
visual and direct representation in which consolidations are made
straightforward.

\CJE{We have intentionally restricted our description to asset/liability relations, ignoring real assets.} We also ignored the representation of private companies, but they can
easily be incorporated as a set of financial assets, real assets and
liabilities, which induce a net worth possessed by the owners. 
Instead we \CJE{focused on} how these tools can be used to
analyze the structure of the state, with its central bank and its
Treasury, and understand the nature of money. We argued that in
sovereign states running sovereign currencies, the Treasury and the
central bank can be meaningfully consolidated, as we showed how their
actions are coordinated thanks to monetary policy. We explained that
the state only \CJE{exogenously} controls the net money through public
deficits, and sets the boundary conditions for the financial structure
of the private sector, but it does not control the amount of credit
which is endogenously determined. Only the salient features have been
illustrated with graphs, but the reader is encouraged to draw the graphs corresponding to any discussion in order to develop graphical representations of monetary systems.

%For Revtex
\ifrevtex

\ifanonymous
\else
  \begin{acknowledgments}
  The author thanks D. D\'efossez-Carme, L. Gastard, M. Lavoie, E. Le
  H\'eron, S. Renaux-Petel, E. Rolet and  J.-P. Uzan for discussions
  and comments, and Johanna Skrzypczyk for assistance with editing the manuscript.
  \end{acknowledgments}
\fi
  
\else
  %For Elsevier
\ifanonymous
\else
  {\it Acknowledgments}: The author thanks D. D\'efossez-Carme,
  L. Gastard, M. Lavoie, E. Le H\'eron, S. Renaux-Petel, E. Rolet and
  J.-P. Uzan for discussions and comments, and Johanna Skrzypczyk for assistance with editing the manuscript.
\fi
\fi

\ifrevtex
\else
\section*{References}
\fi

\end{document}